\documentstyle[epsf]{ar}

\makeatletter

\def\alt{\mathrel{\mathpalette\vereq<}}
\def\vereq#1#2{\lower3pt\vbox{\baselineskip1.5pt \lineskip1.5pt
\ialign{$\m@th#1\hfill##\hfil$\crcr#2\crcr\sim\crcr}}}
\def\agt{\mathrel{\mathpalette\vereq>}}

\newcount\@tempcntc
\def\@citex[#1]#2{\if@filesw\immediate\write\@auxout
{\string\citation{#2}}\fi
  \@tempcnta\z@\@tempcntb\m@ne\def\@citea{}\@cite{\@for\@citeb:=#2\do
    {\@ifundefined
       {b@\@citeb}{\@citeo\@tempcntb\m@ne\@citea
        \def\@citea{,\penalty\@m\ }{\bf ?}\@warning
       {Citation `\@citeb' on page \thepage \space undefined}}%
{\setbox\z@\hbox{\global\@tempcntc0\csname b@\@citeb\endcsname\relax}%
     \ifnum\@tempcntc=\z@ \@citeo\@tempcntb\m@ne
       \@citea\def\@citea{,\penalty\@m}
       \hbox{\csname b@\@citeb\endcsname}%
     \else
      \advance\@tempcntb\@ne
      \ifnum\@tempcntb=\@tempcntc
      \else\advance\@tempcntb\m@ne\@citeo
      \@tempcnta\@tempcntc\@tempcntb\@tempcntc\fi\fi}}\@citeo}{#1}}

\def\@citeo{\ifnum\@tempcnta>\@tempcntb\else\@citea
  \def\@citea{,\penalty\@m}%
  \ifnum\@tempcnta=\@tempcntb\the\@tempcnta\else
   {\advance\@tempcnta\@ne\ifnum\@tempcnta=\@tempcntb \else
\def\@citea{--}\fi
    \advance\@tempcnta\m@ne\the\@tempcnta\@citea\the\@tempcntb}\fi\fi}

\def\tableofcontents{%
    \par\addvspace{24pt plus3pt minus3pt}%
\vfil\vfil
\begin{centering}
{\Large 24 March 1999}\\
\end{centering}

\vfil
\begin{centering}
{\Large Prepared for the}\\
\medskip 
{\Large\it Annual Review of Nuclear and Particle Science}\\
\medskip
{\Large Vol.~49 (1999)}\\ 
\end{centering}
\eject
    \noindent{\normalsize CONTENTS}%
    \vskip6pt
    \@starttoc{toc}}

\makeatother

\begin{document}

\title{Particle Physics from Stars}

\thispagestyle{plain}

\markboth{Georg G.~Raffelt}{Particle Physics from Stars}

\author{Georg G.~Raffelt\affiliation{Max-Planck-Institut f\"ur Physik
(Werner-Heisenberg-Institut)\\ F\"ohringer Ring 6, 80805 M\"unchen,
Germany}}


\begin{abstract}
  Low-mass particles such as neutrinos, axions, other Nambu-Goldstone
  bosons and gravitons are produced in the hot and dense interior of
  stars. Therefore, astrophysical arguments constrain the properties
  of these particles in ways which are often complementary to
  cosmological arguments and to laboratory experiments.  This review
  provides an update on the most important stellar-evolution limits
  and discusses them in the context of other information from
  cosmology and laboratory experiments.
\end{abstract}

\maketitle

%%%%%%%%%%%%%%%%%%%%%%%%%%%%%%%%%%%%%%%%%%%%%%%%%%%%%%%%%%%%%%%%%%%%%%
%% Section 1 %%%%%%%%%%%%%%%%%%%%%%%%%%%%%%%%%%%%%%%%%%%%%%%%%%%%%%%%%
%%%%%%%%%%%%%%%%%%%%%%%%%%%%%%%%%%%%%%%%%%%%%%%%%%%%%%%%%%%%%%%%%%%%%%

\newpage

\section{INTRODUCTION}

Astrophysical and cosmological arguments and observations have become
part of the main-stream methodology to obtain empirical information on
existing or hypothetical elementary particles and their
interactions. The ``heavenly laboratories'' are complementary to
accelerator and non-accelerator experiments, notably at the
``low-energy frontier'' of particle physics, which includes the
physics of neutrinos and other weakly interacting low-mass particles
such as the hypothetical axions, novel long-range forces, and so
forth.

The present review is dedicated to stars as particle-physics
laboratories, or more precisely, to what can be learned about weakly
interacting low-mass particles from the observed properties of stars.
The prime argument is that a hot and dense stellar plasma emits
low-mass weakly interacting particles in great abundance.  They
subsequently escape from the stellar interior directly, without
further interactions, and thus provide a local energy sink for the
stellar medium.  The astronomically observable impact of this
phenomenon provides some of the most powerful limits on the properties
of neutrinos, axions, and the like.

Once the particles have escaped they can decay on their long way to
Earth, allowing one to derive interesting limits on radiative decay
channels from the absence of unexpected x- or $\gamma$-ray fluxes from
the Sun or other stars.

Finally, the weakly interacting particles can be directly detected at
Earth, thus far only the neutrinos from the Sun and supernova (SN)
1987A, allowing one to extract important information on their
properties.

The material covered here has been reviewed in 1990, with a focus on
axion limits, by Turner~\cite{Turner90a} and by
Raffelt~\cite{Raffelt90a}, and a very brief ``Mini-Review'' was
included in the 1998 edition of the Review of Particle
Physics~\cite{Caso98}.  My 1996 book {\it Stars as Laboratories for
Fundamental Physics}~\cite{Raffelt96a} treats these topics in much
greater detail than is possible here.  The present chapter is intended
as a compact, up-to-date, and easily accessible source for the most
important results and methods.

The subject of ``Particle Physics from Stars'' is broader than both my
expertise and the space available here.  I will not touch on the solar
neutrino problem and its oscillation interpretation---this is a topic
unto itself and has been extensively reviewed by other authors, for
example~\cite{Bahcall,Castellani,Bahcall98a}.

The high densities encountered in neutron stars make them ideal for
studies and speculations concerning novel phases of nuclear matter
(e.g.\ meson condensates or quark matter), an area covered by two
recent books~\cite{Glendenning,Weber}. Quark stars are also the
subject of an older review~\cite{Alcock88} and are covered in the
proceedings of two topical conferences~\cite{Madsen92,Vassiliadis95}.

Certain grand unified theories predict the existence of primordial
magnetic monopoles. They would get trapped in stars and then catalyze
the decay of nucleons by the Rubakov-Callan effect. The ensuing
anomalous energy release is constrained by the properties of stars, in
particular neutron stars and white dwarfs, a topic that has been
reviewed a long time ago~\cite{Kolb90a}.  It was re-examined, and the
limits improved, in the wake of the discovery of the faintest white
dwarf ever detected which puts restrictive limits on an anomalous
internal heat source~\cite{Freese98}.

Weakly interacting massive particles (WIMPs), notably in the guise of
the supersymmetric neutralinos, are prime candidates for the cosmic
dark matter. Some of them would get trapped in stars, annihilate with
each other, and produce a secondary flux of high-energy neutrinos. The
search for such fluxes from the Sun and the center of the Earth by
present-day and future neutrino telescopes is the ``indirect method''
to detect galactic particle dark matter, an approach which is
competitive with direct laboratory searches---see~\cite{Jungman96} for
a review.

Returning to the topics which are covered here,
Sections~\ref{sec:sun}--\ref{sec:SN1987A} are devoted to a discussion
of the main stellar objects that have been used to constrain low-mass
particles, viz.\ the Sun, globular-cluster stars, compact stars, and
SN~1987A. In Sections~\ref{sec:neutrinos}--\ref{sec:longrangeforces}
the main constraints on neutrinos, axions, and novel long-range forces
are summarized. Section~\ref{sec:conclusion} is given over to brief
concluding remarks.

%%%%%%%%%%%%%%%%%%%%%%%%%%%%%%%%%%%%%%%%%%%%%%%%%%%%%%%%%%%%%%%%%%%%%%
%% Section 2 %%%%%%%%%%%%%%%%%%%%%%%%%%%%%%%%%%%%%%%%%%%%%%%%%%%%%%%%%
%%%%%%%%%%%%%%%%%%%%%%%%%%%%%%%%%%%%%%%%%%%%%%%%%%%%%%%%%%%%%%%%%%%%%%

\section{THE SUN}
\label{sec:sun}

\subsection{Basic Energy-Loss Argument}
\label{sec:basicargument}

\begin{figure}[h]
\hbox to\hsize{\hss\epsfxsize=3cm\epsfbox{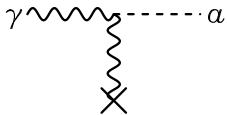}\hss}
\caption{Primakoff production of axions in the Sun.
\label{fig:primakoff}}
\end{figure}

The Sun is the best-known star and thus a natural starting point for
our survey of astrophysical particle laboratories.  It is powered by
hydrogen burning which amounts to the net reaction $4p+2e^-\to{}^4{\rm
He}+2\nu_e+26.73~{\rm MeV}$, giving rise to a measured $\nu_e$ flux
which now provides one of the most convincing indications for neutrino
oscillations~\cite{Bahcall,Castellani,Bahcall98a}.  Instead of
neutrinos from nuclear processes we focus here on particle fluxes
which are produced in thermal plasma reactions.  The photo neutrino
process $\gamma+e^-\to e^-+\nu\bar\nu$ is a case in point, as is the
production of gravitons from electron bremsstrahlung. The solar energy
loss from such standard processes is small, but it may be large for
new particles.  To be specific we consider axions
(Sec.~\ref{sec:axions}) which arise in a variety of reactions, and in
particular by the Primakoff process in which thermal photons mutate
into axions in the electric field of the medium's charged particles
(Fig.~\ref{fig:primakoff}).  In Sec.~\ref{sec:photonlimits} we will
discuss direct search experiments for solar axions, while here we
focus on what is the main topic of this review, the backreaction of a
new energy loss on stars.

The Sun is a normal star which supports itself against gravity by
thermal pressure, as opposed to degenerate stars like white dwarfs
which are supported by electron degeneracy pressure.  If one pictures
the Sun as a self-gravitating monatomic gas in hydrostatic
equilibrium, the ``atoms'' obey the virial theorem $\langle E_{\rm
kin}\rangle=-\frac{1}{2}\, \langle E_{\rm grav}\rangle$. The most
important consequence of this relationship is that extracting energy
from such a system, i.e.~reducing the total energy $\langle E_{\rm
kin}\rangle+\langle E_{\rm grav}\rangle$, leads to contraction and to
an {\it increase\/} of $\langle E_{\rm kin}\rangle$. Therefore, all
else being equal, axion losses lead to contraction and heating.  The
nuclear energy generation rate scales with a high power of the
temperature. Therefore, the heating implied by the new energy loss
causes increased nuclear burning---the star finds a new equilibrium
configuration where the new losses are compensated by an increased
rate of energy generation.

The main lesson is that the new energy loss does not ``cool'' the
star; it leads to heating and an increased consumption of nuclear
fuel. The Sun, where energy is transported from the central nuclear
furnace by radiation, actually overcompensates the losses and
brightens, while it would dim if the energy transfer were by
convection. Either behavior is understood by a powerful ``homology
argument'' where the nonlinear interplay of the equations of stellar
structure is represented in a simple analytic
fashion~\cite{Frieman87}.

The solar luminosity is well measured, yet this brightening effect is
not observable because all else need not be equal.  The present-day
luminosity of the Sun depends on its unknown initial helium mass
fraction $Y$; in a solar model $Y$ has to be adjusted such that
$L_\odot=3.85\times10^{33}~{\rm erg~s^{-1}}$ is reproduced after
$4.6\times10^9~{\rm years}$ of nuclear burning. For solar models with
axion losses the required presolar helium abundance $Y$ as a function
of the axion-photon coupling constant $g_{a\gamma}$ is shown in
Table~\ref{tab:solaraxions}.  The axion luminosity $L_a$ is also given
as well as the central helium abundance $Y_c$, density $\rho_c$, and
temperature $T_c$ of the present-day Sun.

\begin{table}[b]
\centering
\caption{\label{tab:solaraxions} Solar-model parameters and relative
detection rates in the Cl, Ga, and water neutrino observatories as a
function of the axion-photon coupling constant $g_{10}\equiv
g_{a\gamma}/(10^{-10}~{\rm GeV}^{-1})$ according to
Ref.~\protect\cite{Schlattl99}.}
\medskip
\begin{tabular}[8]{cccccccccc}
\noalign{\hrule\vskip2pt}
\noalign{\hrule\vskip3pt}
$g_{10}$ & $L_a$ & $Y$ & $Y_c$ & 
$\rho_c$ & $T_c$ &Cl&Ga&H$_2$O\\
&$[L_\odot]$&&& 
$[{\rm g~cm^{-3}}]$ &$[{\rm 10^7~K}]$\\
\noalign{\vskip3pt\hrule\vskip3pt}
0   &  0    &  0.266 &0.633 & 153.8 & 1.563 
&   1   &   1   &   1  \\
4.5 &  0.04 &  0.265 &0.641 & 158.0 & 1.575 
&   1.07   &   1.16   & 1.20 \\
10  &  0.20 &  0.257 &0.679 & 177.5 & 1.626 
&   1.45   &   2.2  &   2.4 \\
15  &  0.53 &  0.245 &0.751 & 218.3 & 1.722 
&   2.5   &   6.0    & 6.7 \\
20  &  1.21 &  0.228 &0.914 & 324.2 & 1.931 
&   6.4   &   20   &   23  \\
\noalign{\vskip3pt\hrule}
\end{tabular}
\end{table}

Even axion losses as large as $L_\odot$ can be accommodated by
reducing the presolar helium mass fraction from about 27\% to
something like 23\%~\cite{Schlattl99,Raffelt87a}.  The ``standard
Sun'' has completed about half of its hydrogen-burning phase.
Therefore, the anomalous energy losses cannot exceed approximately
$L_\odot$ or else the Sun could not have reached its observed age.
Indeed, for $g_{10}=30$ no consistent present-day Sun could be
constructed for any value of $Y$~\cite{Raffelt87a}.  The emission rate
of other hypothetical particles would have a different temperature and
density dependence than the Primakoff process, yet the general
conclusion remains the same that a novel energy loss must not exceed
approximately~$L_\odot$.

\subsection{Solar Neutrino Measurements}

This crude limit is improved by the solar neutrino flux which has been
measured in five different observatories with three different spectral
response characteristics, i.e.\ by the absorption on chlorine,
gallium, and by the water Cherenkov technique.  The axionic solar
models produce larger neutrino fluxes; in Table~\ref{tab:solaraxions}
we show the expected detection rates for the Cl, Ga, and H$_2$O
experiments relative to the standard case.  For $g_{10}\alt10$ one can
still find oscillation solutions to the observed $\nu_e$ deficit, but
larger energy-loss rates appear to be excluded~\cite{Schlattl99}.

Once the neutrino oscillation hypothesis has been more firmly
established and the mixing parameters are better known, the neutrino
measurements may be used to pin down the central solar temperature,
allowing one to constrain novel energy losses with greater
precision. For now it appears safe to conclude that the Sun does not
emit more than a few tenths of $L_\odot$ in new forms of radiation.

\subsection{Helioseismology}
\label{sec:helioseismology}

Over the past few years the precision measurements of the solar p-mode
frequencies have provided a more reliable way to study the solar
interior.  For example, the convective surface layer is found to reach
down to 0.710--$0.716\,R_\odot$~\cite{Dalsgaard91}, the helium content
of these layers to exceed 0.238~\cite{Scilla97}. Gravitational
settling has reduced the surface helium abundance by about 0.03 so
that the presolar value must have been at least 0.268, in good
agreement with standard solar models. The reduced helium content
required of the axionic solar models in Table~\ref{tab:solaraxions}
disagrees significantly with this lower limit for $g_{10}\geq10$.

\begin{figure}[ht]
\hbox to\hsize{\hss\epsfxsize=9cm\epsfbox{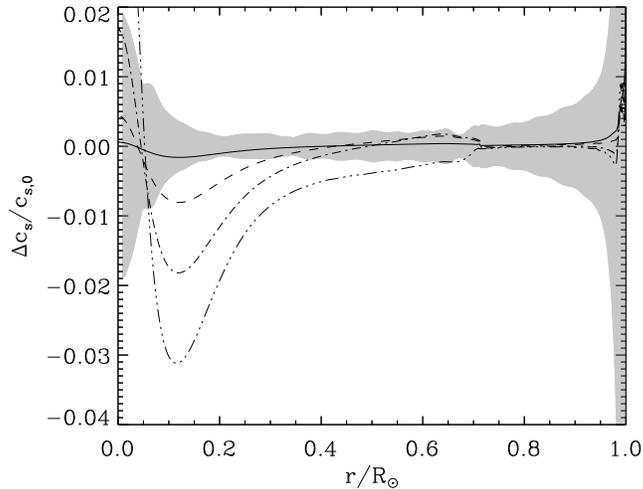}\hss}
\vskip-6pt
\caption{Fractional difference in sound-speed profiles of solar models
with axion losses compared to the reference
model~\protect\cite{Schlattl99}.  The shaded area is the uncertainty
of the seismic model~\protect\cite{Scilla97}.  The axion-photon
coupling constant was $g_{10}$=4.5 (solid line), 10 (short-dashed), 15
(dash-dotted), 20 (dash-dot-dot-dotted).
\label{fig:soundspeed}}
\end{figure}

One may also invert the p-mode measurements to construct a ``seismic
model'' of the solar sound-speed profile, e.g.~\cite{Basu97}.  All
modern standard solar models agree well with the seismic model within
its uncertainties (shaded band in Fig.~\ref{fig:soundspeed}) which
mostly derive from the inversion method itself, not the
measurements. The difference between the sound-speed profile of a
standard solar model and those including axion losses are also shown
in Fig.~\ref{fig:soundspeed}.  For $g_{10}\geq10$ the difference is
larger than the uncertainties of the seismic model, implying a limit
\begin{equation}\label{eq:helioaxionlimit}
g_{a\gamma}\alt10\times10^{-10}~{\rm GeV}^{-1}.
\end{equation}
Other cases may be different in detail, but it appears safe to assume
that any new energy-loss channel must not exceed something like 10\%
of $L_\odot$.

\subsection{``Strongly'' Interacting Particles}
\label{sec:strongcoupling}

Thus far we have assumed that the new particles couple so weakly that
they escape from the stellar interior without further interactions, in
analogy to neutrinos or gravitons.  They emerge from the entire
stellar volume, i.e.~their emission amounts to a local energy sink for
the stellar plasma. But what if the particles interact so strongly
that their mean free path is less than the solar radius?

The impact of such particles on a star compares to that of photons,
which are also ``trapped'' by their ``strong'' interaction. Their
continuous thermal production and re-absorption amounts to the net
transfer of energy from regions of higher temperature to cooler
ones. In the Sun this radiative form of energy transfer is more
important than conduction by electrons or convection, except in the
outer layers. A particle which interacts more weakly than photons is
more effective because it travels a larger distance before
re-absorption---the ability to transfer energy is proportional to the
mean free path. The properties of the Sun roughly confirm the standard
photon opacities, so that a new particle would have to interact more
strongly than photons to be allowed~\cite{Carlson89,Raffelt89a}.

Therefore, contrary to what is sometimes stated in the literature, a
new particle is by no means allowed just because its mean free path is
less than the stellar dimensions. The impact of a new particle is
maximal when its mean free path is of order the stellar radius.  Of
course, usually one is interested in very weakly interacting particles
so that this point is moot.

%%%%%%%%%%%%%%%%%%%%%%%%%%%%%%%%%%%%%%%%%%%%%%%%%%%%%%%%%%%%%%%%%%%%%%
%% Section 3 %%%%%%%%%%%%%%%%%%%%%%%%%%%%%%%%%%%%%%%%%%%%%%%%%%%%%%%%%
%%%%%%%%%%%%%%%%%%%%%%%%%%%%%%%%%%%%%%%%%%%%%%%%%%%%%%%%%%%%%%%%%%%%%%

\section{LIMITS ON STELLAR ENERGY LOSSES}
\label{sec:lowmassstars}

\subsection{Globular-Cluster Stars}
\label{sec:globularcluster}

\subsubsection{Evolution of Low-Mass Stars}
\label{sec:evolution}

The discussion in the previous section suggests that the emission of
new weakly interacting particles from stars primarily modifies the
time scale of evolution. For the Sun this effect is less useful to
constrain particle emission than, say, the modified p-mode frequencies
or the direct measurement of the neutrino fluxes. However, the
observed properties of other stars provide far more restrictive limits
on their evolutionary time scales so that anomalous modes of energy
loss can be far more tightly constrained.  We begin with
globular-cluster stars which, together with SN~1987A, are the most
successful example of astronomical observations that provide
nontrivial limits on the properties of elementary particles.

\begin{figure}[b]
\hbox to\hsize{\hss\epsfxsize=7cm\epsfbox{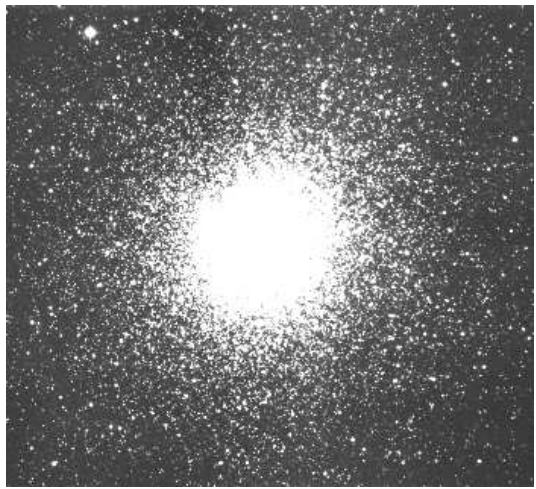}\hss}
\caption{Globular cluster M3.
(Image courtesy of Palomar/Caltech.)
\label{fig:m3}}
\end{figure}

\begin{figure}[ht]
\hbox to\hsize{\hss\epsfxsize=8cm\epsfbox{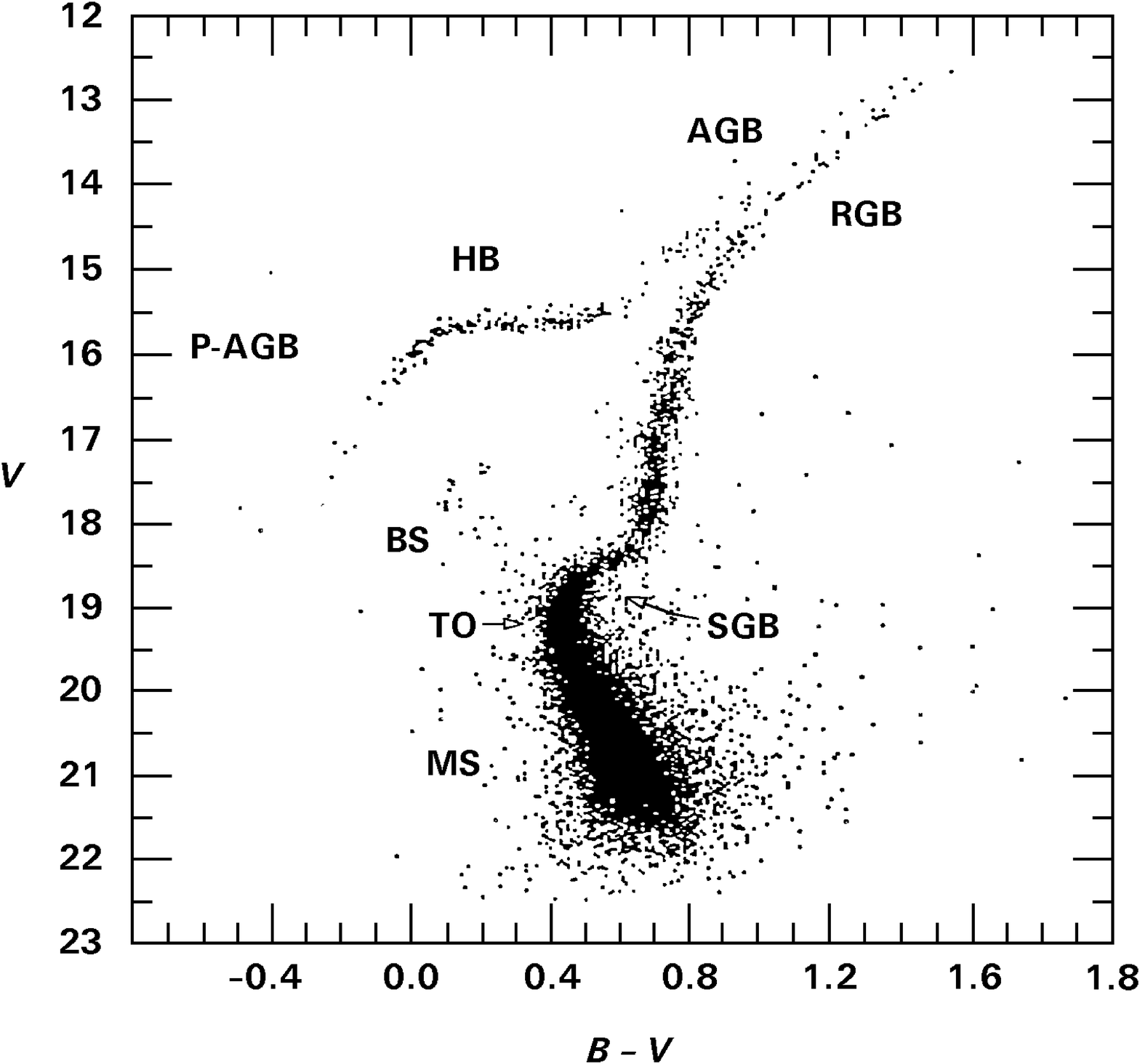}\hss}
\caption{Color-magnitude diagram for the globular cluster M3, based on
the photometric data of 10,637 stars~\protect\cite{Buonanno86}.
Vertically is the brightness in the visual (V) band, horizontally the
difference between B (blue) and V brightness, i.e.\ a measure of the
color and thus surface temperature, where blue (hot) stars lie toward
the left.  The classification for the evolutionary phases is as
follows~\protect\cite{Renzini88}.  MS (main sequence): core hydrogen
burning. BS (blue stragglers). TO (main-sequence turnoff): central
hydrogen is exhausted.  SGB (subgiant branch): hydrogen burning in a
thick shell. RGB (red-giant branch): hydrogen burning in a thin shell
with a growing core until helium ignites. HB (horizontal branch):
helium burning in the core and hydrogen burning in a shell. AGB
(asymptotic giant branch): helium and hydrogen shell burning. P-AGB
(post-asymptotic giant branch): final evolution from the AGB to the
white-dwarf stage.
\label{fig:colmag}}
\end{figure}

Our galaxy has about 150 globular clusters such as M3
(Fig.~\ref{fig:m3}) which are gravitationally bound systems of up to a
million stars. In Fig.~\ref{fig:colmag} the stars of the cluster M3
are arranged according to their color or surface temperature
(horizontal axis) and brightness (vertical axis) in the usual way,
leading to a characteristic pattern which allows for rather precise
tests of the theory of stellar evolution, and notably for quantitative
measurements of certain evolutionary time scales.  Globular clusters
are the oldest objects in the galaxy and thus almost as old as the
universe.  The stars in a given cluster all formed at about the same
time with essentially the same chemical composition, differing
primarily in their mass.  Because more massive stars evolve faster,
present-day globular-cluster stars are somewhat 
below\footnote{The letter ${\cal M}$ denotes stellar masses with
$1\,{\cal M}_\odot=2\times10^{33}~{\rm g}$ the solar mass. The letter
$M$ is traditionally reserved for the absolute stellar brightness (in
magnitudes or mag). The total or bolometric brightness is defined as
$M_{\rm bol}=4.74-2.5\,\log_{10}(L/L_\odot)$, with the solar luminosity
$L_\odot=3.85\times10^{33}~{\rm erg~s^{-1}}$.} $1\,{\cal M}_\odot$ so
that we are concerned with low-mass stars (${\cal M}\alt2\,{\cal
M}_\odot$). Textbook expositions of stellar structure and evolution
are~\cite{Clayton68,Kippenhahn90}.

Stars begin their life on the main sequence (MS) where they burn
hydrogen in their center. Different locations on the MS
in a color-magnitude diagram like Fig.~\ref{fig:colmag} correspond to
different masses, with more massive stars shining more brightly.  When
central hydrogen is exhausted  the star develops a degenerate
helium core, with hydrogen burning in a shell.  Curiously, the stellar
envelope expands, leading to a large surface area and thus a low
surface temperature (red color)---they become ``red giants.''  The
luminosity is governed by the gravitational potential at the edge of
the growing helium core so that these stars become ever brighter: they
ascend the red-giant branch (RGB). The higher a star on the RGB, the
more massive and compact its helium core.

The core grows until about $0.5\,{\cal M}_\odot$ when it has become
dense and hot enough to ignite helium.  The ensuing core expansion
reduces the gravitational potential at its edge and thus lowers the
energy production rate in the hydrogen shell source, dimming these
stars.  Helium ignites at a fixed core mass, but the envelope mass
differs due to varying rates of mass loss on the RGB, leading to
different surface areas and thus surface temperatures.  These stars
thus occupy the horizontal branch (HB) in the color-magnitude
diagram. In Fig.~\ref{fig:colmag} the HB turns down on the left (blue
color) where much of the luminosity falls outside the V filter; in
terms of the total or ``bolometric'' brightness the HB is truly
horizontal.

Finally, when helium is exhausted, a degenerate carbon-oxygen core
develops, leading to a second ascent on what is called the asymptotic
giant branch (AGB). These low-mass stars cannot ignite their
carbon-oxygen core---they become white dwarfs after shedding most of
their envelope.

The advanced evolutionary phases are fast compared with the MS
duration which is about $10^{10}~{\rm yr}$ for stars somewhat below
$1\,{\cal M}_\odot$. For example, the ascent on the upper RGB and the
HB phase each take around $10^8~{\rm yr}$.  Therefore, the
distributions of stars along the RGB and beyond can be taken as an
``isochrone'' for the evolution of a single star, i.e.\ a time-series
of snapshots for the evolution of a single star with a fixed initial
mass.  Put another way, the number distribution of stars along the
different branches are a direct measure for the duration of the
advanced evolutionary phases. The distribution along the MS is
different in that it measures the distribution of initial masses.

\subsubsection{Core Mass at Helium Ignition}
\label{sec:coremass}

Anomalous energy losses modify this picture in measurable ways.  We
first consider an energy-loss mechanism which is more effective in the
degenerate core of a red giant before helium ignition than on the HB
so that the post-RGB evolution is standard.  Since an RGB-star's
helium core is supported by degeneracy pressure there is no feedback
between energy-loss and pressure: the core is actually {\it cooled}.
Helium burning ($3{}^4{\rm He}\to{}^{12}{\rm C}$) depends very
sensitively on temperature and density so that the cooling delays the
ignition of helium, leading to a larger core mass ${\cal M}_c$, with
several observable consequences.

First, the brightness of a red giant depends on its core mass so that
the RGB would extend to larger luminosities, causing an increased
brightness difference $\Delta M_{\rm HB}^{\rm tip}$ between the HB and
the RGB tip.  Second, an increased ${\cal M}_c$ implies an increased
helium-burning core on the HB. For a certain range of colors these
stars are pulsationally unstable and are then called RR Lyrae
stars. From the measured RR Lyrae luminosity and pulsation period one
can infer ${\cal M}_c$ on the basis of their so-called mass-to-light
ratio $A$.  Third, the increased ${\cal M}_c$ increases the luminosity
of RR Lyrae stars so that absolute determinations of their brightness
$M_{\rm RR}$ allow one to constrain the range of possible core masses.
Fourth, the number ratio $R$ of HB stars vs.\ RGB stars brighter than
the HB is modified.

These observables also depend on the measured cluster metallicity as
well as the unknown helium content which is usually expressed in terms
of $Y_{\rm env}$, the envelope helium mass fraction. Since globular
clusters formed shortly after the big bang, their initial helium
content must be close to the primordial value of
\hbox{22--25\%}. $Y_{\rm env}$ should be close to this number because
the initial mass fraction is somewhat depleted by gravitational
settling, and somewhat increased by convective dredge-up of processed,
helium-rich material from the inner parts of the star.

\begin{figure}[b]
\hbox to\hsize{\hss\epsfxsize=\hsize\epsfbox{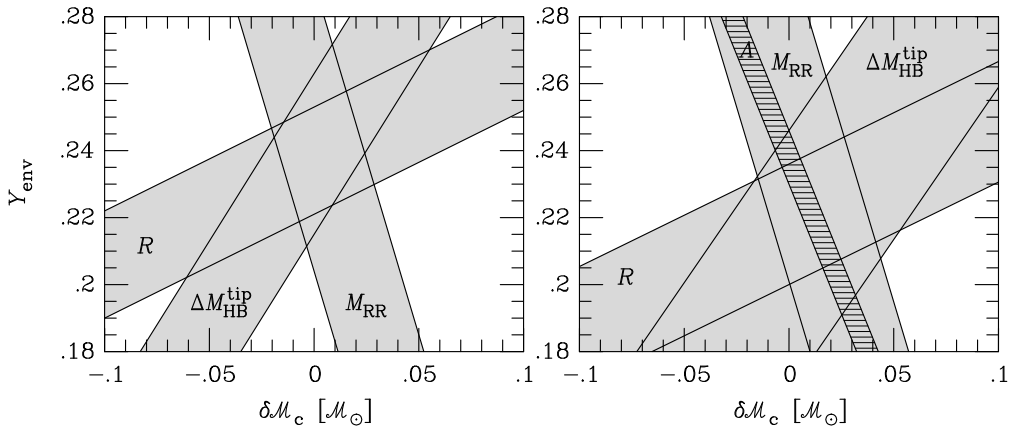}\hss}
\caption{Allowed values for a core-mass excess at helium ignition
$\delta{\cal M}_c$ and the envelope helium mass fraction $Y_{\rm env}$
of evolved globular-cluster stars.  Left
after~\protect\cite{Raffelt96a}, right after~\protect\cite{Catelan96}.
The observables are the brightness difference $\Delta M_{\rm HB}^{\rm
tip}$ between the HB and the RGB tip, the RR Lyrae mass-to-light ratio
$A$, their absolute brightness $M_{\rm RR}$, and the number ratio $R$
between HB and RGB stars.
\label{fig:coremass}}
\end{figure}

An estimate of ${\cal M}_c$ from a global analysis of these
observables except $A$ was performed in~\cite{Raffelt90b} and
re-analysed in~\cite{Raffelt96a}, $A$ was used in~\cite{Castellani93},
and an independent analysis using all four observables
in~\cite{Catelan96}.  In Fig.~\ref{fig:coremass} we show the allowed
core mass excess $\delta {\cal M}_c$ and envelope helium mass fraction
$Y_{\rm env}$ from the analyses~\cite{Raffelt96a,Catelan96};
references to the original observations are found in these papers.

Figure~\ref{fig:coremass} suggests that, within the given
uncertainties, the different observations overlap at the standard core
mass ($\delta{\cal M}_c=0$) and at an envelope helium abundance of
$Y_{\rm env}$ which is compatible with the primordial helium
abundance.  Of course, the error bands do not have a simple
interpretation because they combine observational and estimated
systematic errors, which involve some subjective judgement by the
authors.  The difference between the two panels of
Fig.~\ref{fig:coremass} gives one a sense of how sensitive the
conclusions are to these more arbitrary aspects of the analysis. As a
nominal limit it appears safe to adopt $|\delta{\cal M}_c|\alt 0.025$
or $|\delta{\cal M}_c|/{\cal M}_c\alt 5\%$; how much additional
``safety-margin'' one wishes to include is a somewhat arbitrary
decision which is difficult to make objective in the sense of a
statistical confidence level.

In~\cite{Raffelt96a} it was shown that this limit can be translated
into an approximate limit on the average anomalous energy-loss rate
$\epsilon_x$ of a helium plasma,
\begin{equation}\label{eq:rglimit}
\epsilon_x\alt 10~{\rm erg~g^{-1}~s^{-1}}
\hbox{\quad at\quad} 
T\approx10^8~{\rm K},
\quad
\rho\approx2\times10^5~{\rm g~cm^{-3}}.
\end{equation}
The density represents the approximate average of a red-giant core
before helium ignition; the value at its center is about $10^6~{\rm
g~cm^{-3}}$. The main standard-model neutrino emission process is
plasmon decay $\gamma\to\nu\bar\nu$ with a core average of about
$4~{\rm erg~g^{-1}~s^{-1}}$.  Therefore, Eq.~(\ref{eq:rglimit}) means
that a new energy-loss channel must be less effective than a few times
the standard neutrino losses.

\subsubsection{Helium-Burning Lifetime of Horizontal-Branch Stars}

We now turn to an energy-loss mechanism which becomes effective in a
nondegenerate medium, i.e.\ we imagine that the core expansion after
helium ignition ``switches on'' an energy-loss channel that was
negligible on the RGB. Therefore, the pre-HB evolution is taken to be
standard. As in the case of the Sun (Sec.~\ref{sec:basicargument})
there will be little change in the HB stars' brightness, rather they
will consume their nuclear fuel faster and thus begin to ascend the
AGB sooner. The net observable effect is a reduction of the number of
HB relative to RGB stars.

From the measured HB/RGB number ratios in 15 globular
clusters~\cite{Buzzoni83} and with plausible assumptions about the
uncertainties of other parameters one concludes that the duration of
helium burning agrees with stellar-evolution theory to within about
10\%~\cite{Raffelt96a}.  This implies that the new energy loss of the
helium core should not exceed about 10\% of its standard energy
production rate. Therefore, the new energy-loss rate at average core
conditions is constrained by~\cite{Raffelt96a}
\begin{equation}\label{eq:hblimit}
\epsilon_x\alt 10~{\rm erg~g^{-1}~s^{-1}}
\hbox{\quad at\quad} 
T\approx0.7\times10^8~{\rm K},
\quad
\rho\approx0.6\times10^4~{\rm g~cm^{-3}}.
\end{equation}
This limit is slightly more restrictive than the often-quoted
``red-giant bound,'' corresponding to $\epsilon_x\alt 100~{\rm
erg~g^{-1}~s^{-1}}$ at $T=10^8~{\rm K}$ and $\rho=10^4~{\rm
g~cm^{-3}}$. It was based on the helium-burning lifetime of the
``clump giants'' in open clusters~\cite{Raffelt88d}. They have fewer
stars, leading to statistically less significant limits. The ``clump
giants'' are the physical equivalent of HB stars, except that they
occupy a common location at the base of the RGB, the ``red-giant
clump.''

\subsubsection{Applications}

After the energy-loss argument has been condensed into the simple
criteria of Eqs.~(\ref{eq:rglimit}) and~(\ref{eq:hblimit}) it can be
applied almost mechanically to a variety of cases. The main task is to
identify the dominant emission process for the new particles and to
calculate the energy-loss rate $\epsilon_x$ for a helium plasma at the
conditions specified in Eqs.~(\ref{eq:rglimit})
or~(\ref{eq:hblimit}). The most important limits will be discussed in
the context of specific particle-physics hypotheses in
Secs.~\ref{sec:neutrinos}--\ref{sec:longrangeforces}.  Here we just
mention that these and similar arguments were used to constrain
neutrino electromagnetic
properties~\cite{Raffelt90b,Castellani93,Raffelt88d,%
Sutherland76,Fukugita87,Raffelt89b,Raffelt90c,Raffelt92a},
axions~\cite{Raffelt87a,Dicus78,Mikaelian78,Dicus80,Georgi81,%
Barroso82,Fukugita82a,Fukugita82b,Krauss84,Brodsky86,Pantziris86,%
Raffelt86a,Chanda88,Raffelt90f,Haxton91,Raffelt95a},
paraphotons~\cite{Hoffmann87}, the photo production cross section on
$^4$He of new bosons~\cite{Raffelt88e,Velde89}, the Yukawa couplings
of new bosons to baryons or electrons~\cite{Grifols86a,Grifols89d},
and supersymmetric particles~\cite{Bouquet82,Fukugita82c,Anand84}.

One may also calculate numerical evolution sequences including new
energy losses~\cite{Raffelt87a,Castellani93,Raffelt89b,Raffelt92a,%
Raffelt95a,Sweigart78}. Comparing the results from such studies with
what one finds from Eqs.~(\ref{eq:rglimit}) and~(\ref{eq:hblimit})
reveals that, in view of the overall theoretical and observational
uncertainties, it is indeed enough to use these simple
criteria~\cite{Raffelt96a}.

\subsection{White Dwarfs}

White dwarfs are another case where astronomical observations provide
useful limits on new stellar energy losses. These compact objects are
the remnants of stars with initial masses of up to several~${\cal
M}_\odot$~\cite{Kippenhahn90,Shapiro83}.  For low-mass progenitors the
evolution proceeds as described in Sec.~\ref{sec:evolution}. When they
ascend the asymptotic giant branch they eventually shed most of their
envelope mass. The degenerate carbon-oxygen core, having reached
something like $0.6\,{\cal M}_\odot$, never ignites.  Its subsequent
evolution is simply one of cooling, first dominated by neutrino
losses throughout its volume, later by surface photon emission.

\begin{figure}
\hbox to\hsize{\hss\epsfxsize=6.5cm\epsfbox{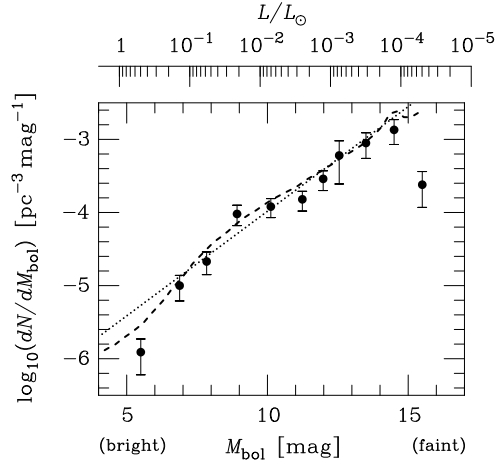}\hss}
\caption{Observed white-dwarf luminosity
function~\protect\cite{Fleming86,Liebert88}.  The dotted line
represents Mestel's cooling law with a constant birthrate of
$10^{-3}~{\rm pc^{-3}~Gyr^{-1}}$.  The dashed line is from the cooling
curve of a $0.6\,{\cal M}_\odot$ white dwarf which includes neutrino
losses~\protect\cite{Koester86}, assuming the same constant birthrate.
\label{fig:wd}}
\end{figure}

The cooling speed can be observationally infered from the ``luminosity
function,'' i.e.\ the white-dwarf number density per brightness
interval. As white dwarfs are intrinsically dim they are observed only
in the solar neighborhood, out to perhaps $100~{\rm pc}$ ($1~{\rm
pc}=3.26~{\rm lyr}$) which is far less than the thickness of the
galactic disk.  The measured luminosity function (Fig.~\ref{fig:wd})
reveals that there are few bright white dwarfs and many faint ones.
The dotted line represents Mestel's cooling
law~\cite{Shapiro83,Mestel52}, an analytic treatment based on surface
photon cooling.  The observed luminosity function dips at the bright
end, a behavior ascribed to neutrino emission which quickly ``switches
off'' as the star cools.

The luminosity function drops sharply at the faint end. Even the
oldest white dwarfs have not yet cooled any further, implying that
they were born 8--12~Gyr ago, in good agreement with the estimated age
of the galaxy. Therefore, a novel cooling agent cannot be much more
effective than the surface photon emission. This conclusion also
follows from the agreement between the implied birthrate with
independent estimates.  The shape of the luminosity function can be
deformed for an appropriate temperature dependence of the particle
emission rate, e.g.\ enhancing the ``neutrino dip'' at the bright end.
Finally, white dwarfs in a certain range of surface temperatures are
pulsationally unstable and are then called ZZ~Ceti stars. The
pulsation period of a few minutes depends on the luminosity, the
period decrease thus on the cooling speed. For G117--B15A the period
change was measured~\cite{Kepler91,Kepler95}, implying a somewhat
large cooling rate. While this discrepancy may be worrisome, probably
these measurements should be taken as an approximate confirmation of
the predicted white-dwarf cooling speed.

White dwarfs were used to constrain the axion-electron
coupling~\cite{Raffelt86b,Nakagawa87,Nakagawa88,Wang92,Blinnikov94,%
Altherr94b}.  It was also noted that the somewhat large period
decrease of G117--B15A could be ascribed to axion
cooling~\cite{Isern92}.  Finally, a limit on the neutrino magnetic
dipole moment was derived~\cite{Blinnikov94}. A detailed review of
these limits is provided in~\cite{Raffelt96a}; they are somewhat
weaker than those from globular-cluster stars, but on the same general
level. Therefore, white-dwarf cooling essentially corroborates some of
the globular-cluster limits, but does not improve on them.

\subsection{Old Neutron Stars}

Neutron stars are the compact remnants of stars with initial masses
beyond about $8\,{\cal M}_\odot$. After their formation in a
core-collapse supernova (Sec.~\ref{sec:SN1987A}) they evolve by
cooling, a process that speeds up by a new energy-loss
channel.  Neutron-star cooling can now be observed by satellite-borne
x-ray measurements of the thermal surface emission of several old
pulsars---a recent review is~\cite{Tsuruta98}.

Limits on axions were derived in~\cite{Iwamoto84,Tsuruta87,Umeda98},
on neutrino magnetic dipole moments in~\cite{Iwamoto95}. These bounds
are much weaker than those from SN~1987A or globular clusters.
Turning this around, anomalous cooling effects by particle emission is
probably not important in old neutron stars, leaving them as
laboratories for many of the other uncertain bits of input physics
such as the existence of new phases of nuclear
matter~\cite{Glendenning,Weber,Tsuruta98,Umeda94a,Umeda94b}.  If a
neutron star converts into a strange-matter star an axion burst
emerges~\cite{Suh98}, but for now this effect has not provided new
empirical information on axion properties.

%%%%%%%%%%%%%%%%%%%%%%%%%%%%%%%%%%%%%%%%%%%%%%%%%%%%%%%%%%%%%%%%%%%%%%
%% Supernova 1987A %%%%%%%%%%%%%%%%%%%%%%%%%%%%%%%%%%%%%%%%%%%%%%%%%%%
%%%%%%%%%%%%%%%%%%%%%%%%%%%%%%%%%%%%%%%%%%%%%%%%%%%%%%%%%%%%%%%%%%%%%%

\section{SUPERNOVAE}
\label{sec:SN1987A}

\subsection{SN~1987A Neutrino Observations}

When the explosion of the star Sanduleak $-69\,202$ was detected on 23
February 1987 in the Large Magellanic Cloud, a satellite galaxy of our
Milky Way at a distance of about 50~kpc (165,000~lyr), it became
possible for the first time to measure the neutrino emission from a
nascent neutron star, turning this supernova (SN~1987A) into one of
the most important stellar particle-physics
laboratories~\cite{Schramm90,Raffelt90d,Koshiba92}.

\begin{figure}[ht]
\hbox to\hsize{\hss\epsfxsize=6.3cm\epsfbox{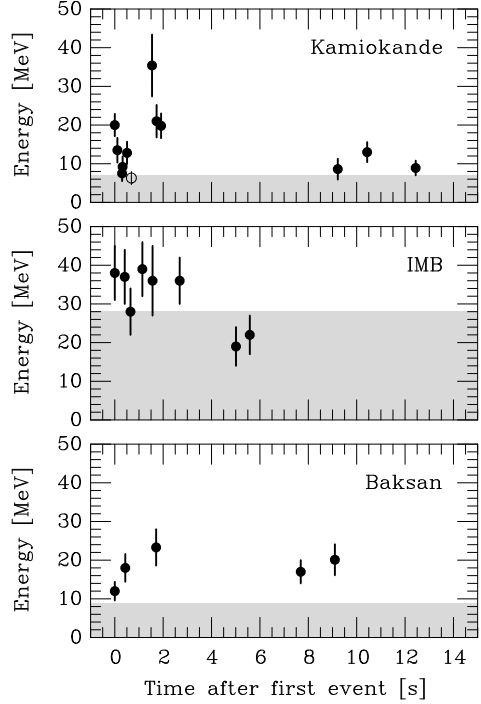}\hss}
\caption{\label{fig:sndat} SN~1987A neutrino observations at
Kamiokande~\protect\cite{Hirata88}, IMB~\protect\cite{Bratton88} and
Baksan~\protect\cite{Alexeyev87}.  The energies refer to the secondary
positrons from the reaction $\bar\nu_e p\to n e^+$.  In the shaded
area the trigger efficiency is less than 30\%. The clocks have unknown
relative offsets; in each case the first event was shifted to
$t=0$. In Kamiokande, the event marked as an open circle is attributed
to background.}
\end{figure}

A type~II supernova explosion
\cite{Brown82,Bethe90,Petschek90,Cooperstein88,Burrows90b,Janka93a} is
physically the implosion of an evolved massive star (${\cal
M}\agt8\,{\cal M}_\odot$) which has become an ``onion-skin structure''
with several burning shells surrounding a degenerate iron core. It
cannot gain further energy by fusion so that it becomes unstable when
it has reached the limiting mass (Chandrasekhar mass) of 1--$2\,{\cal
M}_\odot$ that can be supported by electron degeneracy pressure. The
ensuing collapse is intercepted when the equation of state stiffens at
around nuclear density ($3\times10^{14}~{\rm g~cm^{-3}}$),
corresponding to a core size of a few tens of kilometers.  At
temperatures of tens of MeV this compact object is opaque to
neutrinos. The gravitational binding energy of the newborn neutron
star (``proto neutron star'') of about $3\times10^{53}~{\rm erg}$ is
thus radiated over several seconds from the ``neutrino sphere.''
Crudely put, the collapsed SN core cools by thermal neutrino emission
from its surface.

The neutrino signal from SN~1987A (Fig.~\ref{fig:sndat}) was observed
by the $\bar\nu_e p\to n e^+$ reaction in several
detectors~\cite{Koshiba92}.  The number of events, their energies, and
the distribution over several seconds corresponds well to theoretical
expectations and thus has been taken as a confirmation of the standard
picture that a compact remnant formed which emitted its energy by
quasi-thermal neutrino emission. Detailed statistical analyses of the
data were performed in~\cite{Loredo89,Loredo95}.

The signal does show a number of ``anomalies.'' The average
$\bar\nu_e$ energies infered from the Irvine-Michigan-Brookhaven (IMB)
and Kamiokande observations are quite
different~\cite{Janka89a,Jegerlehner96}. The large time gap of 7.3~s
between the first 8 and the last 3 Kamiokande events looks
worrisome~\cite{Lattimer89}. The distribution of the final-state
positrons from the $\bar\nu_e p\to n e^+$ capture reaction should be
isotropic, but is found to be significantly peaked away from the
direction of the SN~\cite{Velde89,Secco89,Kielczewska90}.  In the
absence of other explanations, these features have been blamed on
statistical fluctuations in the sparse data.

\subsection{Signal Dispersion}
\label{sec:signaldispersion}

A dispersion of the neutrino burst can be caused by a time-of-flight
delay from a nonvanishing neutrino mass~\cite{Zatsepin68}.  The
arrival time from SN~1987A at a distance $D$ would be delayed by
\begin{equation}\label{eq:sndelay}
\Delta t=2.57~{\rm s}\,
\left(\frac{D}{50~{\rm kpc}}\right)\,
\left(\frac{10~{\rm MeV}}{E_\nu}\right)^2\,
\left(\frac{m_\nu}{10~{\rm eV}}\right)^2.
\end{equation}
As the $\bar\nu_e$ were registered within a few seconds and had
energies in the 10~MeV range, $m_{\nu_e}$ is limited to less
than around 10~eV.  Detailed analyses reveal that the pulse duration
is consistently explained by the intrinsic SN cooling time 
and that $m_{\nu_e}\alt 20~{\rm eV}$ is implied as something like a
95\% CL limit~\cite{Loredo89,Kernan95}.

The apparent absence of a time-of-flight dispersion effect of the
$\bar\nu_e$ burst was also used to constrain a ``millicharge'' of
these particles (they would be deflected in the galactic
magnetic field)~\cite{Bahcall,Barbiellini87}, a quantum field theory
with a fundamental length scale~\cite{Fujiwara89}, and deviations from
the Lorentzian rule of adding velocities~\cite{Atzmon94}.  Limits on
new long-range forces acting on the
neutrinos~\cite{Pakvasa89,Grifols88b,Grifols94,Fiorentini89,Malaney95}
seem to be invalidated in the most interesting case of a long-range
leptonic force by screening from the cosmic background
neutrinos~\cite{Dolgov95}.

The SN~1987A observations confirm that the visual SN explosion occurs
several hours after the core-collapse and thus after the neutrino
burst. Again, there is no apparent time-of-flight delay of the
relative arrival times between the neutrino burst and the onset of the
optical light curve, allowing one to confirm the equality of the
relativistic limiting velocity for these particle types to within
$2\times10^{-9}$ \cite{Longo87,Stodolsky88}.  Moreover, the Shapiro
time delay in the gravitational field of the galaxy of neutrinos
agrees with that of photons to within about $4\times10^{-3}$
\cite{Krauss88}, constraining certain alternative theories of
gravity~\cite{Coley88,Almeida89}.

\subsection{Energy-Loss Argument}
\label{sec:energylossargument}

The late events in Kamiokande and IMB reveal that the signal duration
was not anomalously short. Very weakly interacting particles would
freely stream from the inner core, removing energy which otherwise
would
power the late-time neutrino signal. Therefore, its observed duration
can be taken as evidence against such novel cooling effects. This
argument has been advanced to constrain axion-nucleon
couplings~\cite{Ellis87a,Raffelt88a,Turner88a,%
Mayle88,Mayle89,Burrows89a,Burrows90a,Janka96a,Keil97},
majorons~\cite{Grifols88a,Aharonov88a,Aharonov88b,Aharonov89,%
Choi88a,Choi90a,Chang94}, supersymmetric
particles~\cite{Ellis88a,Lau93,Nowakowksi95,%
Grifols89b,Grifols97a,Grifols97b,Grifols98a,Dicus98}, and graviton
emission in quantum-gravity theories with higher
dimensions~\cite{Arkani99,Cullen99}. 
 It has also been used to constrain
right-handed neutrinos interacting by a Dirac
mass term~\cite{Raffelt88a,Gaemers89,Grifols90a,Gandhi90,Mayle93,%
Maalampi91,Turner92a,Pantaleone92a,Burrows92a,Goyal94a,%
Babu92a,Babu92b}, mixed with active
neutrinos~\cite{Kainulainen91a,Raffelt93a}, interacting through
right-handed
currents~\cite{Raffelt88a,Barbieri89a,Grifols90b,Grifols90c,Rizzo91},
a magnetic dipole
moment~\cite{Lattimer88a,Barbieri88a,Goyal94b,Goyal95,Ayala98}, or an
electric form factor~\cite{Mohapatra90a,Grifols89a}.  Many of these
results will be reviewed in
Secs.~\ref{sec:neutrinos}--\ref{sec:longrangeforces} in the context of
specific particle-physics hypotheses.

\begin{figure}[b]
\hbox to\hsize{\hss\epsfxsize=6.3cm\epsfbox{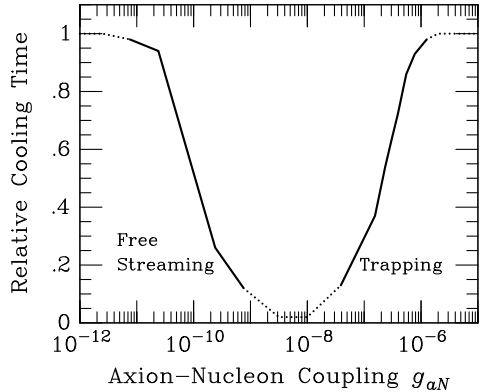}\hss}
\caption{\label{fig:snax}Relative duration of SN neutrino cooling as a
function of the axion-nucleon coupling.  Freely streaming axions are
emitted from the entire core volume, trapped ones from the ``axion
sphere.''  The solid line follows from the numerical
calculations~\protect\cite{Burrows89a,Burrows90a}; the dotted line is
an arbitrary continuation.}
\end{figure}

Here we illustrate the general argument with axions
(Sec.~\ref{sec:axions}) which are produced by nucleon bremsstrahlung
$NN\to NNa$ so that the energy-loss rate depends on the axion-nucleon
Yukawa coupling $g_{aN}$. In Fig.~\ref{fig:snax} we show the expected
neutrino-signal duration as a function of $g_{aN}$. With increasing
$g_{aN}$, corresponding to an increasing energy-loss rate, the signal
duration drops sharply.  For a sufficiently large $g_{aN}$, however,
axions no longer escape freely; they are trapped and thermally emitted
from the ``axion sphere'' at unit optical depth.  Beyond some coupling
strength axions are less important than neutrinos and cannot be
excluded.

However, particles which are on the ``strong interaction'' side of
this argument need not be allowed. They could be important for the
energy-transfer during the infall phase and they could produce events
in the neutrino detectors.  For example, ``strongly coupled'' axions
in a large range of $g_{aN}$ are actually excluded because they would
have produced too many events by their absorption on
$^{16}$O~\cite{Engel90a}.

Likewise, particles on the free-streaming side can cause excess
events in the neutrino detectors. For example, right-handed neutrinos
escaping from the inner core could become ``visible'' by decaying into
left-handed states~\cite{Dodelson92} or by spin-precessing in
the galactic magnetic field if they have a dipole moment.

Returning to the general argument, one can estimate a limit on the
energy-loss rate on the free-streaming side by the simple 
criterion that the new channel should be less effective than the
standard neutrino losses, corresponding to~\cite{Raffelt96a}
\begin{equation}\label{eq:snlimit}
\epsilon_x\alt10^{19}~{\rm erg~g^{-1}~s^{-1}}
\hbox{\quad at\quad}
\rho=3\times10^{14}~{\rm g~cm^{-3}},
\quad
T=30~{\rm MeV}.
\end{equation}
The density is the core average, the temperature an average during the
first few seconds. Some authors find higher temperatures, but for a
conservative limit it is preferable to stick to a value at the lower
end of the plausible range. At these conditions the nucleons are
partially degenerate while the electrons are highly
degenerate. Several detailed numerical studies reveal that this simple
criterion corresponds to approximately halving the neutrino signal
duration~\cite{Raffelt96a}.

A simple analytic treatment is far more difficult on the trapping
side; see~\cite{Turner88a} for an example in the context of axions. 

The SN~1987A energy-loss argument tends to be most powerful at
constraining new particle interactions with nucleons. Therefore, it is
necessary to calculate the interaction rate with a hot and dense
nuclear medium that is dominated by many-body effects. Besides the
sparse data, the theoretical treatment of the emission rate is the
most problematic aspect of this entire method.

\subsection{Radiative Neutrino Decays}
\label{sec:radiativeneutrinodecays}

If neutrinos have masses one expects that the heavier ones are
unstable and decay radiatively as $\nu\to\nu'\gamma$. SN~1987A is
thought to have emitted similar fluxes of neutrinos and antineutrinos
of all flavors so that one would have expected a burst of
$\gamma$-rays in coincidence with the neutrinos.  No excess counts
were observed in the gamma-ray spectrometer (GRS) on the solar maximum
mission (SMM) satellite~\cite{Chupp89,Oberauer93}, leading to
restrictive limits on neutrino 
decays~\cite{Chupp89,Oberauer93,Feilitzsch88,Kolb89a,Bludman92}.  The
GRS happened to go into calibration mode about 223~s after the
neutrino burst, but for low-mass neutrinos ($m_\nu\alt40~{\rm eV}$)
the entire $\gamma$-ray burst would have been captured, leading
to a radiative decay limit of~\cite{Raffelt96a}
\begin{equation}\label{eq:sndecaylimit}
\tau_\gamma/m_\nu\agt0.8\times10^{15}~{\rm s/eV}.
\end{equation}
For higher-mass neutrinos the photon burst would have been stretched
beyond the GRS window. The first $\gamma$-rays from decays near the SN
would arrive in coincidence with the $\bar\nu_e$ burst, but the
$\gamma$-burst duration would be given by something like
Eq.~(\ref{eq:sndelay}).  As a further complication, such higher-mass
neutrinos violate the cosmological mass limit unless they decay
sufficiently fast and thus nonradiatively. Put another way, one must
simultaneously worry about radiative and nonradiative decay
channels---a detailed discussion is in~\cite{Raffelt96a}.

Comparable limits in the higher-mass range were also derived from
$\gamma$-ray data of the Pioneer Venus Orbiter
(PVO)~\cite{Jaffe97}. For $m_\nu\agt0.1~{\rm MeV}$, decay photons
still arrive years after SN~1987A. In 1991 the COMPTEL instrument
aboard the Compton Gamma Ray Observatory looked at the SN~1987A
remnant for about $0.68\times10^6~{\rm s}$, providing the most
restrictive limits in this mass range~\cite{Miller95,Miller96}.

For $m_\nu\agt 2m_e\approx 1.2~{\rm MeV}$, which is only possible for
$\nu_\tau$ with an experimental mass limit of about 18~MeV, the
dominant radiative decay channel is $\nu_\tau\to\nu_e e^+e^-$.  From
SN~1987A one would still expect $\gamma$-rays from the bremsstrahlung
process $\nu_\tau\to\nu_e e^+e^-\gamma$, leading to interesting
limits~\cite{Oberauer93,Jaffe97,Dar87a,Mohapatra94,Schmid98}.

The decay positrons from past galactic SNe would be trapped by the
galactic magnetic fields and thus linger for up to $10^5~{\rm yr}$.
Independently of SN~1987A, measurements of the galactic positron flux
thus provide limits on neutrino decays with final-state
positrons~\cite{Raffelt96a,Dar87b}.

\subsection{Explosion Energetics}
\label{sec:explosionenergetics}

The standard scenario of a type~II SN explosion has it that a shock
wave forms near the edge of the core when its collapse halts at
nuclear density and that this shock wave ejects the mantle of the
progenitor star. However, in typical numerical calculations the shock
wave stalls so that this ``prompt explosion'' scenario does not seem
to work. In the ``delayed explosion'' picture the shock wave is
revived by neutrino heating, perhaps in conjunction with convection,
but even then it appears difficult to obtain a successful or
sufficiently energetic explosion.

Therefore, one may speculate that nonstandard modes of energy transfer
play an important role. An example is Dirac neutrinos with a
magnetic dipole moment of order $10^{-12}\,\mu_{\rm B}$ (Bohr
magnetons). The right-handed (sterile) components would arise in the
deep inner core by helicity-flipping collisions and escape.  They
precess back into interacting states in the large magnetic fields
outside the SN core and heat the shock region; their interaction cross
section would be relatively large because of their large inner-core
energies
\cite{Dar87c,Nussinov87a,Goldman88,Voloshin88,Okun88a,Blinnikov88}.

Certainly it is important not to deposit {\it too much\/} energy in
the mantle and envelope of the star. 99\% of the gravitational binding
energy of the neutron star goes into neutrinos, about 1\% into the
kinetic energy of the explosion, and about 0.01\% into the optical
supernova.  Therefore, neutrinos or other particles emitted from the
core must not decay radiatively within the progenitor's envelope
radius of about 100~s or else too much energy lights
up~\cite{Falk78,Takahara86}.

\subsection{Neutrino Spectra and Neutrino Oscillations}
\label{sec:neutrinooscillations}

Neutrino oscillations can have several interesting ramifications in
the context of SN physics because the temporal and spectral
characteristics of the emission process depend on the neutrino
flavor~\cite{Cooperstein88,Burrows90b,Janka93a,Janka95a}.  The
simplest case is that of the ``prompt $\nu_e$ burst'' which represents
the deleptonization of the outer core layers at about 100~ms after
bounce when the shock wave breaks through the edge of the collapsed
iron core.  This ``deleptonization burst'' propagates through the
mantle and envelope of the progenitor star so that resonant
oscillations take place for a large range of mixing parameters between
$\nu_e$ and some other flavor, notably for most of those values where
the MSW effect operates in the Sun
\cite{Mikheyev86,Arafune87a,Arafune87b,Lagage87,Minakata87,Notzold87,%
Walker87,Kuo88,Minakata88,Rosen88}.  In a water Cherenkov detector
this burst is visible as 
$\nu_e$-$e$ scattering, which is forward peaked,
but one would have expected only a fraction of an event from SN~1987A.
The first event in Kamiokande may be attributed to this signal, but
this interpretation is statistically insignificant.

\begin{figure}[ht]
\hbox to\hsize{\hss\epsfxsize=6.7cm\epsfbox{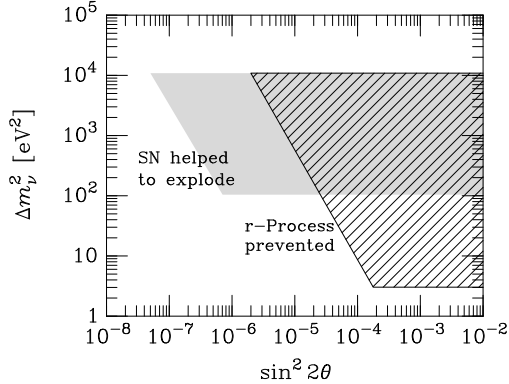}\hss}
\caption{Mass difference and mixing between $\nu_e$ and
$\nu_\mu$ or $\nu_\tau$ where a spectral swap would occur to help
explode supernovae, schematically after~\protect\cite{Fuller92}, and
where it would prevent r-process nucleosynthesis, schematically
after~\protect\cite{Qian93,Qian95,Sigl95a}.
\label{fig:snosci}}
\end{figure}

During the next few hundred milliseconds the shock wave stalls at a few
hundred kilometers above the core and needs rejuvenating.  The
efficiency of neutrino heating can be increased by resonant flavor
oscillations which swap the $\nu_e$ flux with, say, the $\nu_\tau$
one.  Therefore, what passes through the shock wave as a $\nu_e$ was
born as a $\nu_\tau$ at the proto neutron star surface. It has on
average higher energies and thus is more effective at transfering
energy. In Fig.~\ref{fig:snosci} the shaded range of mixing parameters
is where supernovae are helped to explode, assuming a ``normal''
neutrino mass spectrum with $m_{\nu_e}<m_{\nu_\tau}$ \cite{Fuller92}.
Below the shaded region the resonant oscillations take place beyond
the shock wave and thus do not affect the explosion.

The logic of this scenario depends on deviations from strictly thermal
neutrino emission at some blackbody ``neutrino sphere.''  The neutrino
cross sections are very energy dependent and different for different
flavors so that the concept of a neutrino sphere is rather crude---the
spectra are neither thermal nor equal for the different
flavors~\cite{Janka89b,Janka95a}.  The dominant opacity source for
$\nu_e$ is the process $\nu_e+n\to p+e^-$, for $\bar\nu_e$ it is
$\bar\nu_e+p\to n+e^+$, while for $\nu_{\mu,\tau}$ and
$\bar\nu_{\mu,\tau}$ it is neutral-current scattering on nucleons.
Therefore, unit optical depth is at the largest radius (and lowest
medium temperature) for $\nu_e$, and deepest (highest temperature) for
$\nu_{\mu,\tau}$ and $\bar\nu_{\mu,\tau}$.  In typical calculations
one finds a hierarchy $\langle E_{\nu_e}\rangle:\langle
E_{\bar\nu_e}\rangle :\langle E_{\rm others}\rangle\approx
\frac{2}{3}:1:\frac{5}{3}$ with $\langle
E_{\bar\nu_e}\rangle=14$--17~MeV \cite{Janka93a}.  The SN~1987A
observations imply a somewhat lower range of $\langle
E_{\bar\nu_e}\rangle\approx7$--14~MeV~\cite{Loredo89,Jegerlehner96}.

It should be noted that, pending a more detailed numerical
confirmation~\cite{Hardy99}, the difference between the $\bar\nu_e$
and $\nu_{\mu,\tau}$ or $\bar\nu_{\mu,\tau}$ average energies appears
to be smaller than commonly
assumed~\cite{Janka96a,Suzuki91,Suzuki93,Hannestad98}, but there is no
doubt that the $\nu_e$ spectrum is softer than the others. Still, the
quantitative import of flavor oscillations depends on details of the
neutrino spectra formation process in those SN core layers where the
diffusion approximation for the neutrino transport is no longer valid,
yet neutrinos are still trapped.

A few seconds after core bounce the shock wave has long since taken
off, leaving behind a relatively dilute ``hot bubble'' above the
neutron-star surface. This region is one suspected site for r-process
heavy-element synthesis, which requires a neutron-rich environment
\cite{Woosley92,Meyer92,Witti94,Takahashi94,Meyer94,Meyer95,Meyer98}.
The neutron-to-proton ratio, which is governed by the $\beta$
reactions $\nu_e+n\to p+e^-$ and $\bar\nu_e+p\to n+e^+$, is shifted to
a neutron-rich phase if $\langle E_{\nu_e}\rangle<\langle
E_{\bar\nu_e}\rangle$ as for standard neutrino spectra.  Resonant
oscillations can again swap the $\nu_e$ flux with another one,
inverting this hierarchy of energies.  In the hatched range of mixing
parameters shown in Fig.~\ref{fig:snosci} the \hbox{r-process} would
be disturbed \cite{Qian93,Qian95,Sigl95a,Pantaleone95}.  On the other
hand, $\nu_e\to\nu_s$ oscillations into a sterile neutrino could
actually help the r-process by removing some of the neutron-stealing
$\nu_e$ \cite{Laughlin99,Nunokawa97b}.

A large body of recent literature was devoted to explaining the large
kick velocities of the observed radio pulsars as a ``neutrino rocket
effect.'' The required few-percent anisotropy of the SN neutrino
emission was attributed to an intricate interplay between the
magnetic-field induced neutrino dispersion relation and resonant
oscillations
\cite{Kusenko96,Qian97,Kusenko97a,Kusenko97b,Akhmedov97b,Grasso98,%
Horvat98,Kusenko98}. However, due to a conceptual error the effect was
vastly overestimated~\cite{Janka99} so that the pulsar kicks do not
seem to be related to neutrino oscillations in any obvious way.

If the mixing angle between $\nu_e$ and some other flavor is large,
the $\bar\nu_e$ flux from a SN contains a significant fraction of
oscillated states that were born as $\bar\nu_\mu$ or $\bar\nu_\tau$
and thus should have higher average energies. The measured SN~1987A
event energies are already somewhat low, a problem so strongly
exacerbated by oscillations that a large-mixing-angle solution of the
solar neutrino deficit poses a
problem~\cite{Jegerlehner96,Kernan95,Smirnov94}.  This conclusion,
however, depends on the standard predictions for the average neutrino
energies which may not hold up to closer scrutiny as mentioned above.

%%%%%%%%%%%%%%%%%%%%%%%%%%%%%%%%%%%%%%%%%%%%%%%%%%%%%%%%%%%%%%%%%%%%%%
%% Section 5 %%%%%%%%%%%%%%%%%%%%%%%%%%%%%%%%%%%%%%%%%%%%%%%%%%%%%%%%%
%%%%%%%%%%%%%%%%%%%%%%%%%%%%%%%%%%%%%%%%%%%%%%%%%%%%%%%%%%%%%%%%%%%%%%

\section{LIMITS ON NEUTRINO PROPERTIES}
\label{sec:neutrinos}

\subsection{Masses and Mixing}

Astrophysics and cosmology play a fundamental role for neutrino
physics as the properties of stars and the universe at large provide
some of the most restrictive limits on nonstandard properties of these
elusive particles.  Therefore, it behoves us to summarize what the
astrophysical arguments introduced in the previous sections
teach us about neutrinos.

Unfortunately, stars do not tell us very much about neutrino masses,
the holy grail of neutrino physics.  The current
discourse~\cite{Valle98,Kayser98,Smirnov99} centers on the
interpretation of the solar~\cite{Bahcall98a} and
atmospheric~\cite{Fukuda98} neutrino anomalies and the LSND
experiment~\cite{LSND96,LSND98} which all provide very suggestive
evidence for neutrino oscillations. Solar neutrinos imply a $\Delta
m_\nu^2$ of about $10^{-5}~{\rm eV}^2$ (MSW solutions) or
$10^{-10}~{\rm eV}^2$ (vacuum oscillations), atmospheric neutrinos
$10^{-3}$--$10^{-2}~{\rm eV}^2$, and the LSND experiment 0.3--$8~{\rm
eV}^2$.  Taken together, these results require a fourth flavor, a
sterile neutrino, which is perhaps the most spectacular implication of
these experiments, but of course also the least secure.

Core-collapse supernovae appear to be the only case in stellar
astrophysics, apart from the solar neutrino flux, where neutrino
oscillations can be important. However, Fig.~\ref{fig:snosci} reveals
that the experimentally favored mass differences negate a role of
neutrino oscillations for the explosion mechanism or r-process
nucleosynthesis, except perhaps when sterile neutrinos
exist~\cite{Laughlin99,Nunokawa97b}.  Oscillations affect the
interpretation of the SN~1987A
signal~\cite{Jegerlehner96,Kernan95,Smirnov94} and that of a future
galactic SN~\cite{Qian94,Choubey98,Fuller98}.  However, as discussed
in Sec.~\ref{sec:neutrinooscillations}, the main challenge is to
develop a quantitatively more accurate understanding of supernovae as
neutrino sources before relying on 
relatively fine points of the neutrino
spectral characteristics to learn about neutrino mixing
parameters.

Oscillation experiments reveal only mass {\it differences\/} so that
one still needs to worry about the absolute neutrino mass scale. The
absence of anomalous SN~1987A signal dispersion 
(Sec.~\ref{sec:signaldispersion}) gives a 
limit~\cite{Loredo89,Kernan95}
\begin{equation}
m_{\nu_e}\alt 20~{\rm eV},
\end{equation}
somewhat weaker than current laboratory bounds.  A high-statistics
observation of a galactic SN by a detector like Superkamiokande could
improve this limit to about $3~{\rm eV}$ by using the fast
rise-time of the neutrino burst as a measure of dispersion
effects~\cite{Totani98}. If the neutrino mass differences are indeed
very small, this limit carries over to the other flavors. One can
derive an independent mass limit on $\nu_\mu$ and $\nu_\tau$ in the
range of a few $10~{\rm eV}$ if one identifies a neutral-current
signature in a water Cherenkov
detector~\cite{Seckel91,Krauss92,Fiorentini97a,Beacom98a,Beacom98b},
or if a future neutral-current
detector provides an additional measurement~\cite{Cline94,Smith97}.

The SN~1987A energy-loss argument (Sec.~\ref{sec:energylossargument})
provides a limit on a neutrino Dirac mass of
\cite{Raffelt96a,Raffelt88a,Gaemers89,Grifols90a,Gandhi90,Mayle93,%
Burrows92a}
\begin{equation} 
m_\nu({\rm Dirac})\alt30~{\rm keV}.
\end{equation}
It is based on the idea that trapped Dirac neutrinos produce their
sterile component with a probability of about $(m_\nu/2E_\nu)^2$ in
collisions and thus feed energy into an invisible channel.  This 
result was important 
in the discourse on Simpson's 17~keV neutrino
which is now only of historical interest~\cite{Morrison93}.

\subsection{Dipole and Transition Moments}

\subsubsection{Electromagnetic Form Factors}

Neutrino electromagnetic interactions would provide for a great
variety of astrophysical implications.  In the vacuum, the most general
neutrino interaction with the electromagnetic field
is~\cite{Mohapatra91,Winter91}
\begin{equation}
{\cal L}_{\rm int}=-F_1\bar\psi\gamma_\mu\psi A^\mu
-G_1\bar\psi\gamma_\mu\gamma_5\psi\partial_\mu F^{\mu\nu}
-{\textstyle\frac{1}{2}}
\bar\psi\sigma_{\mu\nu}(F_2+G_2\gamma_5)\psi F^{\mu\nu},
\end{equation}
where $\psi$ is the neutrino field, $A^\mu$ the electromagnetic vector
potential, and $F^{\mu\nu}$ the field-strength tensor.  The form
factors are functions of $Q^2$ with $Q$ the energy-momentum transfer.
In the $Q^2\to0$ limit $F_1$ is the electric 
charge, $G_1$ an anapole moment, $F_2$ a magnetic, and $G_2$ an
electric dipole moment.

If neutrinos are electrically strictly neutral, corresponding to
$F_1(0)=0$, they still have a charge radius, usually defined as
$\langle r^2\rangle=6 \partial F_1(Q^2)/e\partial Q^2|_{Q^2=0}$. This
form factor provides for a contact interaction, not for a long-range
force, and as such modifies processes with $Z^0$
exchange~\cite{Lucio85,Auriemma87,Degrassi89,Musolf91,Gongora92}.  As
astrophysics provides no precision test for the effective strength of
neutral-current interactions, this form factor is best probed in
laboratory experiments~\cite{Salati94}. Likewise, the anapole
interaction vanishes in the $Q^2\to0$ limit and thus represents a
modification to the standard neutral-current interaction, with no
apparent astrophysical consequences.

The most interesting possibility are magnetic and electric dipole and
transition moments.  If the standard model is extended to include
neutrino Dirac masses, the magnetic dipole moment is
$\mu_\nu=3.20\times10^{-19}\,\mu_{\rm B}\,m_\nu/{\rm eV}$ where
$\mu_{\rm B}=e/2m_e$ is the Bohr magneton~\cite{Mohapatra91,Winter91}.
An electric dipole moment $\epsilon_\nu$ violates CP, and both are
forbidden for Majorana neutrinos.  Including flavor mixing implies
electric and magnetic transition moments for both Dirac and Majorana
neutrinos, but they are even smaller due to GIM cancellation.  These
values are far too small to be of any experimental or astrophysical
interest. Significant neutrino electromagnetic form factors
require a more radical extension of the standard model, for example
the existence of right-handed currents.

\subsubsection{Plasmon Decay in Stars}

\begin{figure}[b]
\hbox to\hsize{\hss\epsfxsize=8cm\epsfbox{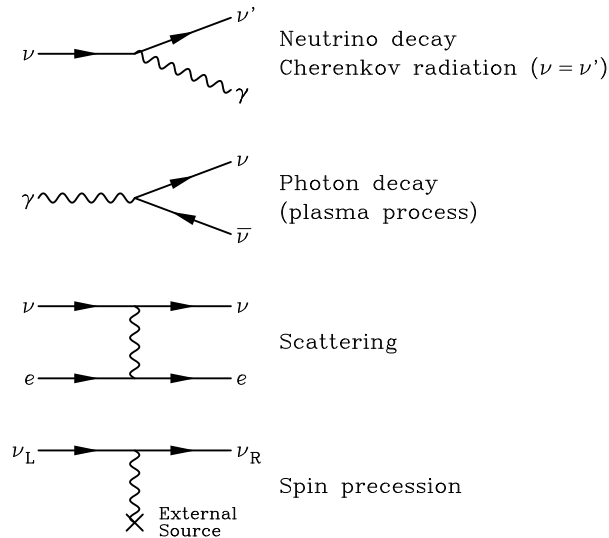}\hss}
\caption{Processes with neutrino electromagnetic
dipole or transition moments.\label{fig:processes}}
\end{figure}

Dipole or transition moments allow for several interesting processes
(Fig.~\ref{fig:processes}).  For the purpose of deriving limits, the
most important case is $\gamma\to\nu\bar\nu$ which is kinematically
possible in a plasma because the photon acquires a dispersion relation
which roughly amounts to an effective mass.  Even without anomalous
couplings, the plasmon decay proceeds because the charged particles of
the medium induce an effective neutrino-photon interaction. Put
another way, even standard neutrinos have nonvanishing electromagnetic
form factors in a medium~\cite{DOlivo89,Altherr94a}.  The standard
plasma process~\cite{Adams63,Zaidi65,Haft94} dominates the neutrino
production in white dwarfs or the cores of globular-cluster red
giants.

The plasma process was first used in~\cite{Bernstein63} to constrain
neutrino electromagnetic couplings. Numerical
implementations of the nonstandard rates 
in stellar-evolution calculations 
are~\cite{Castellani93,Raffelt89b,Raffelt92a,Blinnikov94}.
The helium-ignition argument in globular clusters
(Sec.~\ref{sec:coremass}), equivalent to 
Eq.~(\ref{eq:rglimit}),
implies a limit~\cite{Raffelt96a,Raffelt90b,Raffelt90c,Raffelt92a}
\begin{equation}\label{eq:dipolelimit}
\mu_\nu\alt3\times10^{-12}\,\mu_{\rm B},
\end{equation}
applicable to magnetic and electric dipole and transition moments for
Dirac and Majorana neutrinos.  Of course, the final-state neutrinos
must be lighter than the photon plasma mass which is around 10~keV for
the relevant conditions.

The corresponding laboratory limits are much weaker~\cite{Caso98}.
The most restrictive bound is $\mu_{\nu_e}<1.8\times10^{-10}\,\mu_{\rm
B}$ at 90\% CL from a measurement of the $\bar\nu_e$-$e$-scattering
cross section involving reactor sources.  A significant improvement
should become possible with the MUNU experiment~\cite{Broggini99}, but
it is unlikely that the globular-cluster limit can be reached anytime
soon.

\subsubsection{Radiative Decay}

\begin{figure}[ht]
\hbox to\hsize{\hss\epsfxsize=7cm\epsfbox{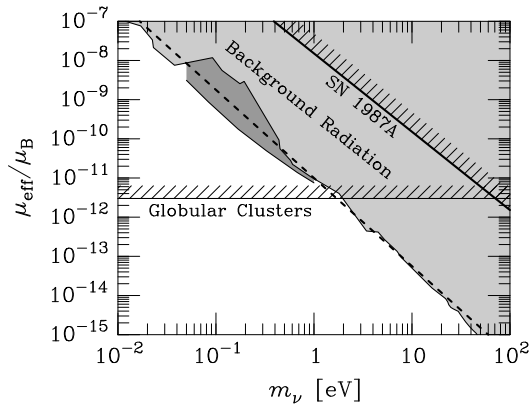}\hss}
\caption{Astrophysical limits on neutrino dipole moments. The
light-shaded background-radiation limits are
from~\protect\cite{Ressell90}, the dark-shaded ones
from~\protect\cite{Biller98,Raffelt98}, the dashed line is the
approximation formula in Eq.~(\protect\ref{eq:munulimits}),
bottom line.
\label{fig:munu}}
\end{figure}

A neutrino mass eigenstate $\nu_i$ may decay to another one $\nu_j$ by
the emission of a photon, where the only contributing form factors are
the magnetic and electric transition moments. The inverse radiative
lifetime is found to be~\cite{Mohapatra91,Winter91}
\begin{equation}\label{eq:radiativedecay}
\tau_\gamma^{-1}=\frac{|\mu_{ij}|^2+|\epsilon_{ij}|^2}{8\pi}
\left(\frac{m_i^2-m_j^2}{m_i}\right)^3
=5.308~{\rm s}^{-1}
\left(\frac{\mu_{\rm eff}}{\mu_{\rm B}}\right)^2
\left(\frac{m_i^2-m_j^2}{m_i^2}\right)^3
\left(\frac{m_i}{{\rm eV}}\right)^3,
\end{equation}
where $\mu_{ij}$ and $\epsilon_{ij}$ are the transition moments while
$|\mu_{\rm eff}|^2\equiv|\mu_{ij}|^2+|\epsilon_{ij}|^2$.  Radiative
neutrino decays have been constrained from the absence of decay
photons of reactor $\bar\nu_e$ fluxes~\cite{Oberauer87}, the solar
$\nu_e$ flux~\cite{Cowsik77,Raffelt85}, and the SN~1987A neutrino
burst~\cite{Chupp89,Oberauer93,Feilitzsch88,Kolb89a,Bludman92}.  For
$m_\nu\equiv m_i\gg m_j$ these limits can be expressed as
\begin{equation}\label{eq:munulimits}
\frac{\mu_{\rm eff}}{\mu_{\rm B}}\;\alt\;
\cases{\hbox to1.8cm{$0.9{\times}10^{-1}$\hfil}({\rm eV}/m_\nu)^2
&Reactor ($\bar\nu_e$),\cr
\hbox to1.8cm{$0.5{\times}10^{-5}$\hfil}({\rm eV}/m_\nu)^2
&Sun ($\nu_e$),\cr
\hbox to1.8cm{$1.5{\times}10^{-8}$\hfil}({\rm eV}/m_\nu)^2
&SN~1987A (all flavors),\cr
\hbox to1.8cm{$1.0{\times}10^{-11}$\hfil}({\rm eV}/m_\nu)^{9/4}
&Cosmic background (all flavors).\cr}
\end{equation}
In this form the SN~1987A limit
applies for $m_\nu\alt 40~{\rm eV}$
as explained in Sec.~\ref{sec:radiativeneutrinodecays}. 
The decay of cosmic background neutrinos would contribute to the
diffuse photon backgrounds, excluding the shaded areas in
Fig.~\ref{fig:munu}. They are approximately delineated by the dashed
line, corresponding to the bottom line in
Eq.~(\ref{eq:munulimits}). More restrictive limits obtain for certain
masses above 3~eV from the absence of emission features from
several galaxy clusters~\cite{Henry81,Davidsen91,Bershady91}. 

For low-mass neutrinos the $m_\nu^3$ phase-space factor in
Eq.~(\ref{eq:radiativedecay}) is so punishing that the
globular-cluster limit is the most restrictive one for $m_\nu$ below a
few eV. This is precisely the mass range which today appears favored
from neutrino oscillation experiments.  Turning this around, the
globular-cluster limit implies that radiative decays of low-mass
neutrinos do not seem to have observable consequences.

For masses above about 30~eV one must invoke fast invisible decays in
order to avoid a conflict with the cosmological mass limit. In this
case radiative decay limits involve the total lifetime as another
parameter; the SN~1987A limits have been interpreted in this sense
in~\cite{Raffelt96a,Kolb89a,Jaffe97,Miller96}.

\subsubsection{Cherenkov Effect}

Another form of ``radiative decay'' is the Cherenkov effect
$\nu\to\nu+\gamma$, which involves the same initial- and final-state
neutrino.  This process is kinematically allowed for photons with
$\omega^2-{\bf k}^2<0$, which obtains in certain media or in external
magnetic fields.  The neutrino may have an anomalous dipole moment,
but there is also a standard-model photon coupling induced by the
medium or the external field.  Thus far it does not look as if the
neutrino Cherenkov effect has any strong astrophysical significance
(see~\cite{Ioannissyan97} for a review of the literature).

\subsubsection{Spin-Flip Scattering}

The magnetic or electric dipole interaction couples neutrino fields of
opposite chirality. In the relativistic limit this implies that a
neutrino flips its helicity in an ``electromagnetic collision,'' which
in the Dirac case produces the sterile component.  The active states
are trapped in a SN core so that spin-flip collisions open an
energy-loss channel in the form of sterile states. Conversely, the
SN~1987A energy-loss argument (Sec.~\ref{sec:energylossargument})
allows one to derive a limit~\cite{Barbieri88a,Ayala98},
\begin{equation}
\mu_\nu({\rm Dirac})\alt3\times10^{-12}\,\mu_{\rm B},
\end{equation}
for both electric and magnetic dipole and transition moments.  It is
the same as the globular-cluster limit Eq.~(\ref{eq:dipolelimit}),
which however includes the Majorana case.

Spin-flip collisions would also populate the sterile Dirac components
in the early universe and thus increase the effective number of
thermally excited neutrino degrees of freedom.  Full thermal
equilibrium is attained for $\mu_\nu({\rm
Dirac})\agt60\times10^{-12}\,\mu_{\rm B}$ \cite{Fukugita87,Elmfors97}.
In view of the SN~1987A and globular-cluster limits this result
assures us that big-bang nucleosynthesis remains undisturbed.

\subsubsection{Spin and Spin-Flavor Precession}

Neutrinos with magnetic or electric dipole moments spin-precess in
external magnetic fields~\cite{Fujikawa80,Okun86a}.  For example,
solar neutrinos can precess into sterile and thus undetectable states
in the Sun's magnetic
field~\cite{Werntz70,Cisneros71,Voloshin86c}. The same for SN
neutrinos in the galactic magnetic field where an important effect
obtains for $\mu_\nu\agt10^{-12}\,\mu_{\rm B}$. Moreover, the
high-energy sterile states emitted by spin-flip collisions from the
inner SN core could precess back into active ones and cause events
with anomalously high energies in SN neutrino detectors, an effect
which probably requires $\mu_\nu({\rm Dirac})\alt10^{-12}\,\mu_{\rm
B}$ from the SN~1987A signal~\cite{Barbieri88a,Notzold88}.  For the
same general $\mu_\nu$ magnitude one may expect an anomalous rate of
energy transfer to the shock wave in a SN, helping with the explosion
(Sec.~\ref{sec:explosionenergetics}).

In a medium the refractive energy shift for active neutrinos relative
to sterile ones creates a barrier to the spin
precession~\cite{Voloshin86a,Voloshin86b}. The mass difference has the
same effect if the precession is between different flavors through a
transition moment~\cite{Schechter81}.  However, the mass and
refractive terms may cancel, leading to resonant spin-flavor
oscillations in the spirit of the MSW
effect~\cite{Akhmedov88a,Akhmedov88b,Barbieri88,Lim88}.  This
mechanism can explain all solar neutrino
data~\cite{Akhmedov95,Guzzo98}, but requires rather large toroidal
magnetic fields in the Sun since the neutrino magnetic (transition)
moments have to obey the globular-cluster limit of
Eq.~(\ref{eq:dipolelimit}).  For Majorana neutrinos, the spin-flavor
precession amounts to transitions between neutrinos and antineutrinos
so that the observation of anti\-neutrinos from the Sun would be a
diagnostic for this effect~\cite{Barbieri91,Fiorentini97b,Pastor98}.

Large magnetic fields exist in SN cores so that spin-flavor precession
could play an important role there, with possible consequences for the
explosion mechanism, r-process nucleosynthesis, or the measurable
neutrino
signal~\cite{Athar95,Totani96,Akhmedov97,Bruggen97,Nunokawa97}.  The
downside of this richness of phenomena is that there are so many
unknown parameters (electromagnetic neutrino properties, masses,
mixing angles) as well as the unknown magnetic field strength and
distribution that it is difficult to come up with reliable limits or
requirements on neutrino properties.  The SN phenomenon is probably
too complicated to serve as a laboratory to pin down electromagnetic
neutrino properties, but it clearly is an environment where these
properties could have far-reaching consequences.

\subsection{Millicharged Particles}

It is conceivable that neutrinos carry small electric charges if
charge conservation is not exact~\cite{Babu90,Maruno91} or if the
families are not sequential~\cite{Takasugi92,Babu92c,Foot93}.
Moreover, new particles with small electric charges are motivated in
certain models with a ``mirror sector'' and a slightly broken mirror
symmetry~\cite{Holdom86}.  Therefore, it is interesting to study the
experimental, astrophysical, and cosmological bounds on
``millicharged''
particles~\cite{Dobroliubov90,Davidson91,Babu92d,Mohapatra92,%
Davidson94}.

\begin{figure}[b]
\hbox to\hsize{\hss\epsfxsize=7cm\epsfbox{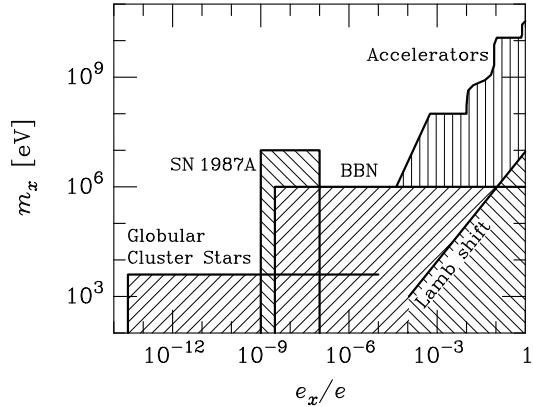}\hss}
\caption{Limits on the electric charge $e_x$ and mass $m_x$
of generic millicharged particles which may be neutrinos or new
particles.  In order to avoid overclosing the universe, additional
model-dependent parameter regions are excluded. The big-bang
nucleosynthesis (BBN) excluded region is larger in some models.
\label{fig:milli}}
\end{figure}

A model-independent $\nu_e$ charge limit arises from the absence of
dispersion of the SN~1987A neutrino signal in the galactic magnetic
field~\cite{Bahcall,Barbiellini87}
\begin{equation}
e_{\nu_e}\alt3\times10^{-17}\,e.
\end{equation}
If charge conservation holds in neutron decay, $e_{\nu_e}\alt
3\times10^{-21}\,e$ results, based on a limit for the neutron charge
of $e_n=(-0.4\pm1.1)\times10^{-21}e$~\cite{Baumann88} and on the
neutrality of matter which was found to be
$e_p+e_e=(0.8\pm0.8)\times10^{-21}e$~\cite{Marinelli84}.  The measured
$\nu_\mu$-$e$ cross section implies
$e_{\nu_\mu}\alt10^{-9}\,e$~\cite{Babu92d}.

Generic millicharged particles (charge $e_x$, mass $m_x$) could appear
as virtual states and would thus modify the Lamb shift unless
$e_x<0.11 e\,m_x/{\rm MeV}$~\cite{Davidson91}.  A number of limits
follow from a host of previous accelerator
experiments~\cite{Davidson91} and a recent dedicated search at
SLAC~\cite{Prinz98}---see Fig.~\ref{fig:milli}.

Millicharged particles are produced by the plasmon decay process and
thus drain energy from stars.  In globular clusters, the emission rate
is almost the same for HB stars and red giants before helium ignition,
in contrast with the magnetic-dipole case. Therefore,
Eqs.~(\ref{eq:rglimit}) and~(\ref{eq:hblimit}) give an almost
identical limit~\cite{Haft94}
\begin{equation}
e_x\alt2\times10^{-14}\,e,
\end{equation}
applicable for $m_x$ below a few keV.  The SN~1987A energy-loss
argument extends the exclusion range to about 10~MeV for
$10^{-9}\,e\alt e_x\alt10^{-7}\,e$ \cite{Raffelt96a,Mohapatra90a}.

The usual big-bang nucleosynthesis (BBN) limit on the effective number
of neutrino species $N_{\rm eff}$ 
provides another constraint.  A millicharged
neutrino is of Dirac nature so that its right-handed component adds
one effective species.  If the millicharged particles are not
neutrinos, then depending on their spin 
$N_{\rm eff}$ may increase even more.
If BBN excludes one extra species  one finds
$e_x\alt3\times10^{-9}\,e$ for $m_x\alt1~{\rm MeV}$~\cite{Davidson91}.
More stringent limits apply in certain models where the millicharged
particles are associated with a shadow sector~\cite{Davidson94}.

Further regions in Fig.~\ref{fig:milli} are excluded to avoid
``overclosing'' the universe by the new
particles~\cite{Davidson91,Davidson94}.  However, because their relic
density depends on their annihilation cross section, it is necessary
to specify a model.  It is hard to imagine new particles which
interact solely through their small electric charge!

\subsection{Nonstandard Weak Interactions}

\subsubsection{Right-Handed Currents}

Right-handed (r.h.) weak interactions may exist on some level, e.g.\
in left-right symmetric models where the r.h.\ gauge bosons differ
from the standard ones by their mass. In the low-energy limit relevant
for stars one may account for the new couplings by a r.h.\ Fermi
constant $\epsilon G_{\rm F}$ where $\epsilon$ is a small
dimensionless parameter.  In left-right symmetric models one finds
explicitly for charged-current processes $\epsilon^2_{\rm
CC}=\zeta^2+[m({W_L})/m({W_R})]^2$ where $m(W_{L,R})$ are the l.h.\
and r.h.\ gauge boson masses and $\zeta$ is the left-right mixing
parameter~\cite{Barbieri89a}.

Assuming that neutrinos are Dirac particles, a SN core loses energy
into r.h.\ states as an ``invisible channel'' by the process $e+p\to
n+\nu_{e,R}$. The SN~1987A energy-loss argument
(Sec.~\ref{sec:energylossargument}) then requires $\epsilon_{\rm
CC}\alt10^{-5}$ \cite{Raffelt96a,Raffelt88a,Barbieri89a}. Laboratory
experiments yield a weaker limit of order $\epsilon_{\rm
CC}\alt3\times10^{-2}$ \cite{Jodidio86}, but do not depend on the
assumed existence of r.h.\ neutrinos.

For neutral currents the dominant emission process is $NN\to
NN\nu_R\bar\nu_R$ which is subject to saturation effects as in the
case of axion emission~\cite{Janka96a}. One then finds $\epsilon_{\rm
NC}\alt3\times10^{-3}$ \cite{Raffelt96a}, somewhat less restrictive
than the original limits of~\cite{Raffelt88a,Barbieri89a}; see also
\cite{Grifols90b,Grifols90c,Rizzo91}. This bound is also somewhat less
restrictive than $\epsilon_{\rm NC}\alt10^{-3}$ found from big-bang
nucleosynthesis~\cite{Ellis86}.

\subsubsection{Secret Neutrino Interactions and Majorons}

The neutrino-neutrino cross section is not known experimentally.  It
could be anomalously large if neutrino Majorana masses were to arise
from a suitable majoron
model~\cite{Gelmini81,Chicashige81,Berezhiani92,Kikuchi94}.
``Secret'' neutrino-neutrino interactions were constrained by the fact
that the SN~1987A neutrino signal was not depleted by collisions with
cosmic background neutrinos~\cite{Kolb87}. Supernova physics with
majorons and SN~1987A limits were discussed
in~\cite{Grifols88a,Aharonov88a,Aharonov88b,Aharonov89,Choi88a,%
Choi90a,Chang94,Kolb82,Dicus83,Manohar87,Konoplich88,Fuller88,%
Berezhiani89}. There is little doubt that majoron models will have an
important impact on SN physics for neutrino-majoron
Yukawa couplings in
the $10^{-6}$--$10^{-3}$ range. The existing literature, however, is
too confusing for this author to come up with a clear synthesis of
what SN physics implies for majoron models.

\subsubsection{Flavor-Changing Neutral Currents}

In certain models the neutrino neutral current has an
effective flavor-changing component. Neutrinos propagating in matter
then have medium-induced mixings and thus can oscillate even if they
are strictly massless~\cite{Valle87,Langacker88}.  Naturally, this
phenomenon can be important for the oscillation of
solar~\cite{Guzzo91,Roulet91,Barger91,Bergmann98} and
supernova~\cite{Nunokawa96,Bergmann99} neutrinos.

%%%%%%%%%%%%%%%%%%%%%%%%%%%%%%%%%%%%%%%%%%%%%%%%%%%%%%%%%%%%%%%%%%%%%%
%% Section 6 %%%%%%%%%%%%%%%%%%%%%%%%%%%%%%%%%%%%%%%%%%%%%%%%%%%%%%%%%
%%%%%%%%%%%%%%%%%%%%%%%%%%%%%%%%%%%%%%%%%%%%%%%%%%%%%%%%%%%%%%%%%%%%%%

\section{AXIONS AND OTHER PSEUDOSCALARS}
\label{sec:axions}

\subsection{Interaction Structure}
\label{ssec:axioninteractions}

New spontaneously broken global symmetries imply the existence of
Nambu-Goldstone bosons that are massless and as such present the most
natural case (besides neutrinos) for using stars as particle-physics
laboratories.  Massless scalars would lead to new long-range forces
(Sect.~\ref{sec:longrangeforces}) so that we may focus here on
pseudoscalars. The most prominent example are axions which were
proposed more than twenty years ago as a solution to the strong CP
problem~\cite{PecceiQuinn77a,PecceiQuinn77b,Weinberg78,Wilczek78}; for
reviews see~\cite{Kim87,Cheng88} and for the latest developments the
proceedings of a topical conference~\cite{Sikivie99}. We use axions as
a generic example---it will be obvious how to extend the following
results and discussions to other cases.

Actually, axions are only ``pseudo Nambu-Goldstone bosons'' in that
the spontaneously broken chiral Peccei-Quinn symmetry $U_{\rm PQ}(1)$
is also explicitly broken, providing these particles with a small mass
\begin{equation}\label{eq:axionmass}
m_a=0.60~{\rm eV}~\frac{10^7~{\rm GeV}}{f_a}.
\end{equation}
Here, $f_a$ is the Peccei-Quinn scale, an energy scale which is
related to the vacuum expectation value of the field that breaks
$U_{\rm PQ}(1)$.  The properties of Nambu-Goldstone bosons are always
related to such a scale which is the main quantity to be constrained
by astrophysical arguments, while Eq.~(\ref{eq:axionmass}) is specific
to axions and allows one to express limits on $f_a$ in terms of $m_a$.

In order to calculate the axionic energy-loss rate from stellar
plasmas one needs to specify the interaction with the medium
constituents. The interaction with a fermion $j$
(mass $m_j$) is generically
\begin{equation}\label{eq:axionfermioncoupling}
{\cal L}_{\rm int}=\frac{C_j}{2f_a}\,
\bar\Psi_j\gamma^\mu\gamma_5\Psi_j\partial_\mu a
\hbox{\quad or\quad}
-i\,\frac{C_j m_j}{f_a}\,\bar\Psi_j\gamma_5\Psi_ja,
\end{equation}
where $\Psi_j$ is the fermion and $a$ the axion field and $C_j$ is a
model-dependent coefficient of order unity.
The combination $g_{aj}\equiv C_jm_j/f_a$
plays the role of a Yukawa coupling and $\alpha_{aj}\equiv
g_{aj}^2/4\pi$ acts as an ``axionic fine structure constant.'' The
derivative form of the interaction is more fundamental in that it is
invariant under $a\to a+a_0$ and thus respects the Nambu-Goldstone
nature of these particles. The pseudoscalar form is usually
equivalent, but one has to be careful when calculating processes where
two Nambu-Goldstone bosons are attached to one fermion line, for
example an axion and a pion attached to a
nucleon~\cite{Raffelt88a,Carena89,Choi89,Turner89a,Iwamoto89}.

The dimensionless couplings $C_i$ depend on the detailed
implementation of the Peccei-Quinn mechanism. Limiting our discussion
to ``invisible axion models'' where $f_a$ is much larger than the
scale of electroweak symmetry breaking, it is conventional to
distinguish between models of the DFSZ type (Dine, Fischler,
Srednicki~\cite{Dine81}, Zhitnitski\u\i~\cite{Zhitniskii80}) and of
the KSVZ type (Kim~\cite{Kim79}, Shifman, Vainshtein,
Zakharov~\cite{Shifman80}). In KSVZ models, axions have no tree-level
couplings to the standard quarks or leptons, yet axions couple to
nucleons by their generic mixing with the neutral pion. The latest
analysis gives numerically~\cite{Keil97}
\begin{equation}\label{eq:ksvzcouplings}
C_p=-0.34,\qquad C_n=0.01
\end{equation}
with a statistical uncertainty of about $\pm0.04$ and an
estimated systematic uncertainty of roughly the same magnitude.  The
tree-level couplings to standard quarks and leptons in the DFSZ model
depend on an angle $\beta$ which measures the ratio of vacuum
expectation values of two Higgs fields. One finds~\cite{Keil97}
\begin{equation}\label{eq:dfszcouplings}
C_e={\textstyle\frac{1}{3}}\,\cos^2\beta,\quad
C_p=-0.07-0.46\,\cos^2\beta,\quad 
C_n=-0.15+0.38\,\cos^2\beta,
\end{equation}
with similar uncertainties as in the KSVZ case.

The CP-conserving interaction between photons and pseudoscalars is
commonly expressed in terms of an inverse energy scale $g_{a\gamma}$
according to
\begin{equation}\label{eq:axionphotoncoupling}
{\cal L}_{\rm int}={\textstyle\frac{1}{4}} g_{a\gamma}
F_{\mu\nu}\tilde F^{\mu\nu}a=- 
g_{a\gamma}{\bf E}\cdot{\bf B}\,a,
\end{equation}
where $F$ is the electromagnetic field-strength tensor and $\tilde F$
its dual. For axions
\begin{equation}\label{eq:cgamma}
g_{a\gamma}=\frac{\alpha}{2\pi f_a}\,C_\gamma,\qquad
C_\gamma=\frac{E}{N}-1.92\pm0.08,
\end{equation}
where $E/N$ is the ratio of the electromagnetic and over color
anomalies, a model-dependent ratio of small integers.  In the DFSZ
model or grand unified models one has $E/N=8/3$, for which
$C_\gamma\approx0.75$, but one can also construct models with $E/N=2$,
which significantly reduces the axion-photon
coupling~\cite{Kaplan85}. The value of $C_\gamma$ in a great variety
of cases was reviewed in~\cite{Cheng95,Kim98}.

\subsection{Limits on the Interaction Strength}

\subsubsection{Photons}
\label{sec:photonlimits}

The axion interaction with fermions or photons allows for numerous
reactions which can produce axions in stars, which may imply limits on
the axion coupling strength. Beginning with photons, pseudoscalars
interact according to the Lagrangian of
Eq.~(\ref{eq:axionphotoncoupling}) which allows for the decay
$a\to2\gamma$. In stellar plasmas the photon-axion interaction also
makes possible the Primakoff conversion $\gamma\leftrightarrow a$ in
the electric fields of electrons and nuclei~\cite{Dicus78}---see
Fig.~\ref{fig:primakoff}.  For low-mass pseudoscalars the emission
rate was calculated for various degrees of electron degeneracy
in~\cite{Raffelt86a,Altherr94b,Raffelt88b}, superseding an earlier
calculation where screening effects had been
ignored~\cite{Fukugita82a}.

\begin{figure}[b]
\hbox to\hsize{\hss\epsfxsize=10cm\epsfbox{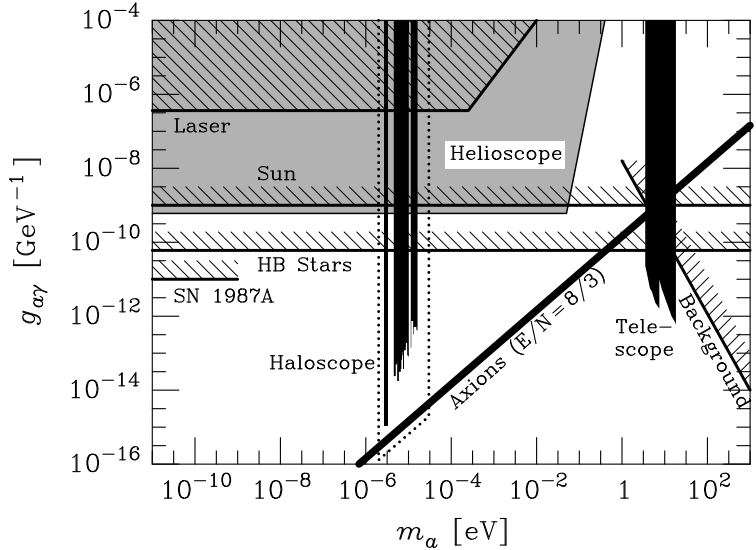}\hss}
\caption{Limits to the axion-photon coupling $g_{a\gamma}$ as defined
in Eq.~(\protect\ref{eq:axionphotoncoupling}).  They apply to any
pseudoscalar except for the ``haloscope'' search which assumes that
these particles are the galactic dark matter; the dotted region marks
the projected sensitivity range of the ongoing dark-matter axion
searches. For higher masses than shown here the pertinent limits are
reviewed in~\protect\cite{Masso97}.
\label{fig:axgam}}
\end{figure}

The helioseismological constraint on solar energy losses then leads to
Eq.~(\ref{eq:helioaxionlimit}) as a bound on $g_{a\gamma}$.
Figure~\ref{fig:axgam} shows this constraint (``Sun'') in the context
of other bounds; similar plots are found
in~\cite{Caso98,Raffelt96a,Masso95a,Masso97,Cameron93}.  For axions
the relationship between $g_{a\gamma}$ and $m_a$ is indicated by the
heavy solid line, assuming $E/N=8/3$.

One may also search directly for solar axions.  One method
(``helioscope'') is to direct a dipole magnet toward the Sun, allowing
solar axions to mutate into \hbox{x-rays}
 by the inverse Primakoff
process~\cite{Sikivie83,Bibber89}.  A pilot experiment was not
sensitive enough~\cite{Lazarus92}, but the exposure time was
significantly increased in a new experiment in Tokyo where a dipole
magnet was gimballed like a telescope so that it could follow the
Sun~\cite{Moriyama98a,Moriyama98b}.  The resulting limit
$g_{a\gamma}\alt 6\times10^{-10}~{\rm GeV}^{-1}$ is more restrictive
than Eq.~(\ref{eq:helioaxionlimit}).  Another helioscope project was
begun in Novosibirsk several years ago~\cite{Vorobev95}, but its
current status has not been reported for some time.
An intruiging project (SATAN) at CERN would
use a decommissioned LHC test magnet that could be mounted on a
turning platform to achieve reasonable periods of alignment with the
Sun~\cite{Zioutas98}.  This setup could begin to compete with the
globular-cluster limit of Eq.~(\ref{eq:globularaxionlimit}).

The axion-photon transition in a macroscopic magnetic field is
analogous to neutrino oscillations and thus depends on the particle
masses~\cite{Raffelt88}. For a large mass difference the transition is
suppressed by the momentum mismatch of particles with equal energies.
Therefore, the Tokyo limit applies only for $m_a\alt0.03~{\rm eV}$. In
a next step one will fill the helioscope with a pressurized gas,
giving the photon a dispersive mass to overcome the momentum mismatch.

An alternative method is ``Bragg diffraction,'' which uses uses the
strong electric field of a crystal lattice which has large Fourier
components for the required momentum
transfer~\cite{Buchmuller90,Paschos94,Creswick98}.  The experiment has
been performed using Ge detectors which were originally built to
search for neutrinoless double-beta decay and for WIMP dark matter;
the crystal serves simultaneously as a Primakoff ``transition agent''
and as an x-ray detector. A first limit of the SOLAX
Experiment~\cite{Solax98} of $g_{a\gamma}\alt27\times10^{-10}~{\rm
GeV}^{-1}$ is not yet compatible with Eq.~(\ref{eq:helioaxionlimit})
and thus not self-consistent.  In the future one may reach this limit,
but prospects to go much further appear dim~\cite{Cebrian99}.

The Primakoff conversion of stellar axions can also proceed in the
magnetic fields of Sun spots or in the galactic magnetic field so that
one might expect anomalous x- or $\gamma$-ray fluxes from the
Sun~\cite{Carlson96}, the red supergiant Betelgeuse~\cite{Carlson95},
or SN~1987A~\cite{Brockway96,Grifols96}.  Observations of SN~1987A
yield $g_{a\gamma}\alt0.1\times10^{-10}~{\rm GeV}^{-1}$ for nearly
massless pseudoscalars with $m_a\alt10^{-9}~{\rm eV}$. A similar limit
obtains from the isotropy of the cosmic x-ray background which would
be modified by the conversion to axions in the galactic magnetic
field~\cite{Krasnikov96}. Axion-photon conversion in the magnetic
fields of stars, the galaxy, or the early universe were also studied
in~\cite{Raffelt88,Morris86,Yoshimura88,Yanagida88,%
Gnedin92,Gnedin94,Gnedin96,Gnedin97,Carlson94}, but no additional
limits emerged.

The existence of massless pseudoscalars would cause a photon
birefringence effect in pulsar magnetospheres, leading to a
differential time delay between photons of opposite helicity and thus
to $g_{a\gamma}\alt0.5\times10^{-10}~{\rm GeV}^{-1}$~\cite{Mohanty93}.

A laser beam in a laboratory magnetic field would also be subject to
vacuum birefringence~\cite{Maiani86}, adding to the QED Cotton-Mouton
effect. First pilot experiments~\cite{Cameron93,Semertzidis90} did not
reach the QED level.  Two vastly improved current projects are
expected to get there~\cite{Bakalov98,Lee95}, but they will stay far
away from the ``axion line'' in Fig.~\ref{fig:axgam}. With a laser
beam in a strong magnet one can also search for Primakoff axion
production and subsequent back-conversion, but a pilot experiment
naturally did not have the requisite sensitivity~\cite{Ruoso92}.  The
exclusion range of current laser experiments is schematically
indicated in Fig.~\ref{fig:axgam}.

\begin{figure}[b]
\hbox to\hsize{\hss\epsfxsize=8cm\epsfbox{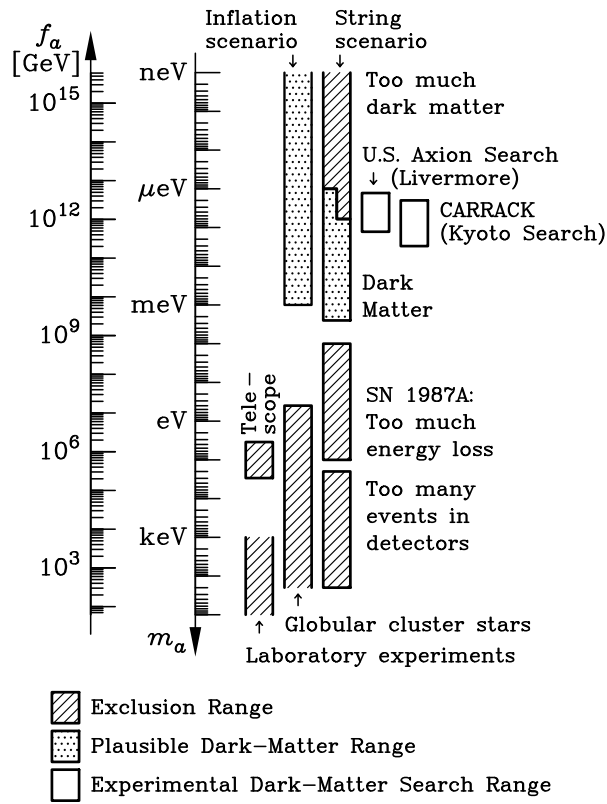}\hss}
\caption{Astrophysical and cosmological exclusion regions (hatched)
for the axion mass $m_a$, or equivalently the Peccei-Quinn scale
$f_a$.  The globular-cluster limit depends on the axion-photon
coupling; it was assumed that $E/N=8/3$ as in GUT models or the DFSZ
model.  The SN~1987A limits depend on the axion-nucleon couplings; the
shown case corresponds to the KSVZ model and approximately to the DFSZ
model.  The dotted ``inclusion regions'' indicate where axions could
plausibly be the cosmic dark matter.  Most of the allowed range in the
inflation scenario requires fine-tuned initial conditions.  In the
string scenario the plausible dark-matter range is somewhat
controversial as indicated by the step in the low-mass end of the
``inclusion bar.''  Also shown is the projected sensitivity range of
the search experiments for galactic dark-matter axions.
\label{fig:axsum}}
\end{figure}

The most important limit on the photon coupling of pseudoscalars
derives from the helium-burning lifetime of HB stars in globular
clusters, i.e.\ from Eq.~(\ref{eq:hblimit}),
\begin{equation}\label{eq:globularaxionlimit}
g_{a\gamma}\alt 0.6\times10^{-10}~{\rm GeV}^{-1}.
\end{equation}
For $m_a\agt10~{\rm keV}$ this limit quickly degrades as the emission
is suppressed when the particle mass exceeds the stellar temperature.
For a fixed temperature, the Primakoff energy-loss rate decreases with
increasing density so that Eq.~(\ref{eq:rglimit}) implies a less
restrictive constraint.  Equation~(\ref{eq:globularaxionlimit}) was
first stated in~\cite{Raffelt96a}, superseding the slightly less
restrictive but often-quoted ``red-giant bound''
of~\cite{Raffelt87a}---see the discussion after
Eq.~(\ref{eq:hblimit}).  The axion relation
Eq.~(\ref{eq:cgamma}) leads to
\begin{equation}
m_a C_\gamma\alt0.3~{\rm eV}
\hbox{\quad and\quad}
f_a/C_\gamma\agt 2\times10^7~{\rm GeV}.
\end{equation}
In the DFSZ model and grand unified models, $C_\gamma\approx 0.75$ so
that $m_a\alt0.4~{\rm eV}$ and $f_a\agt1.5\times10^7~{\rm GeV}$
(Fig.~\ref{fig:axsum}).  For models in which $E/N=2$ and thus
$C_\gamma$ is very small, the bounds are significantly weaker.

On the basis of their two-photon coupling alone, pseudoscalars can
reach thermal equilibrium in the early universe. Their subsequent
$a\to2\gamma$ decays would contribute to the cosmic photon
backgrounds~\cite{Masso97,Mori96}, excluding a non-trivial
$m_a$-$g_{a\gamma}$-range (Fig.~\ref{fig:axgam}).  Some of the
pseudoscalars would end up in galaxies and clusters of galaxies.
Their decay would produce an optical line feature that was not
found~\cite{Bershady91,Ressell91,Overduin93}, leading to the
``telescope'' limits in Fig.~\ref{fig:axgam}.  For axions, the
telescope limits exclude an approximate mass range 4--14~eV even for a
small $C_\gamma$.

Axions with a mass in the $\mu{\rm eV}$ ($10^{-6}~{\rm eV}$) range
could be the dark matter of the universe (Sec.~\ref{sec:cosmology}).
The Primakoff conversion in a microwave cavity placed in a
strong magnetic field (``haloscope'') allows one to search for
galactic dark-matter axions~\cite{Sikivie83}. Two pilot
experiments~\cite{Wuensch89,Hagmann90} and first results from a
full-scale search~\cite{Hagmann98} already exclude a range of coupling
strength shown in Fig.~\ref{fig:axgam}. The new generation of
full-scale experiments~\cite{Sikivie99,Hagmann98,Ogawa96,Yamamoto98}
should cover the dotted area in Fig.~\ref{fig:axgam}, perhaps leading
to the discovery of axion dark matter.

\subsubsection{Electrons}

Pseudoscalars which couple to electrons are produced
by the Compton process $\gamma+e^-\to e^-+a$
\cite{Mikaelian78,Fukugita82b,Brodsky86,Raffelt86a,Chanda88,%
Raffelt95a} and by the electron bremsstrahlung process 
$e^-+ (A,Z)\to(A,Z)+e^-+a$
\cite{Krauss84,Raffelt86a,Raffelt90f,Nakagawa87,Nakagawa88,%
Altherr94b,Iwamoto84}.
A standard solar model yields an axion luminosity
of~\cite{Raffelt86a} $L_a=\alpha_{ae}\,6.0\times10^{21}\,L_\odot$
where $\alpha_{ae}$ is the axion electron ``fine-structure constant''
as defined after Eq.~(\ref{eq:axionfermioncoupling}).  The
helioseismological constraint $L_a\alt0.1\,L_\odot$ of
Sec.~\ref{sec:helioseismology} implies
$\alpha_{ae}\alt2\times10^{-23}$. White-dwarf
cooling gives~\cite{Raffelt96a,Raffelt86b}
$\alpha_{ae}\alt1.0\times10^{-26}$, while the most
restrictive limit is from the delay of helium ignition in low-mass 
red-giants~\cite{Raffelt95a} in the spirit of Eq.~(\ref{eq:rglimit})
\begin{equation}\label{eq:axionelectronlimit}
\alpha_{ae}\alt0.5\times10^{-26}
\hbox{\quad or\quad}
g_{ae}\alt2.5\times10^{-13}.
\end{equation}
For $m_a\agt T\approx 10~{\rm keV}$ this limit quickly degrades
because the emission from a thermal plasma is suppressed.
With Eq.~(\ref{eq:axionfermioncoupling}) one finds for axions
\begin{equation}
m_a C_e\alt0.003~{\rm eV}
\hbox{\quad and\quad}
f_a/C_e\agt 2\times10^9~{\rm GeV}.
\end{equation}
In KSVZ-type models $C_e=0$ at tree level so that no interesting limit
obtains. In the DFSZ model $m_a\cos^2\beta\alt0.01~{\rm eV}$ and
$f_a/\cos^2\beta\agt0.7\times10^9~{\rm GeV}$.  Since $\cos^2\beta$ can
be very small, there is no generic limit on $m_a$.

\subsubsection{Nucleons}

The axion-nucleon coupling strength is primarily constrained by the
SN~1987A energy-loss argument~\cite{Ellis87a,Raffelt88a,Turner88a,%
Mayle88,Mayle89,Burrows89a,Burrows90a,Janka96a,Keil97}.  The main
problem is to estimate the axion emission rate reliably. In the early
papers it was based on a somewhat naive calculation of the
bremsstrahlung process $NN\to NNa$, using quasi-free nucleons that
interact perturbatively through a one-pion exchange potential.
Assuming an equal axion coupling $g_{aN}$ to protons and neutrons this
treatment leads to the \hbox{$g_{aN}$-dependent} shortening of the
SN~1987A neutrino burst of Fig.~\ref{fig:snax}. However, in a dense
medium the bremsstrahlung process likely saturates, reducing the naive
emission rate by as much as an order of magnitude~\cite{Janka96a}.
With this correction, and assuming that the neutrino burst was not
shortened by more than half, one reads from Fig.~\ref{fig:snax} an
excluded range
\begin{equation}\label{eq:snaxlimit}
3\times10^{-10}\alt g_{aN}\alt3\times10^{-7}.
\end{equation}
With Eq.~(\ref{eq:axionfermioncoupling}) this implies an exclusion
range
\begin{equation}
0.002~{\rm eV}\alt m_aC_N\alt2~{\rm eV}
\hbox{\quad and\quad}
3\times10^6~{\rm GeV}\alt f_a/C_N\alt3\times10^9~{\rm GeV}.
\end{equation}
For KSVZ axions the coupling to neutrons disappears while
$C_p\approx-0.34$. With a proton fraction of about 0.3 one estimates
an effective $C_N\approx0.2$ so that~\cite{Raffelt96a,Janka96a}
\begin{equation}
0.01~{\rm eV}\alt m_a\alt10~{\rm eV}
\hbox{\quad and\quad}
0.6\times10^6~{\rm GeV}\alt f_a\alt0.6\times10^9~{\rm GeV}
\end{equation}
is excluded.  

In a detailed numerical study the values for $C_n$ and $C_p$
appropriate for the KSVZ model and for the DFSZ model with different
choices of $\cos^2\beta$ were implemented~\cite{Keil97}.  For KSVZ
axions one finds a limit $m_a\alt0.008~{\rm eV}$, while it varies
between about 0.004 and $0.012~{\rm eV}$ for DFSZ axions, depending on
$\cos^2\beta$. In view of the large overall uncertainties it is
probably good enough to remember $m_a\alt 0.01~{\rm eV}$ as a generic
limit (Fig.~\ref{fig:axsum}).

Axions on the ``strong interaction side'' of the exclusion range
Eq.~(\ref{eq:snaxlimit}) would have produced excess counts in the
neutrino detectors by their absorption on oxygen if $1\times10^{-6}\alt
g_{aN}\alt1\times10^{-3}$ \cite{Engel90a}.  For KSVZ axions this
crudely translates into $20~{\rm eV}\alt m_a\alt 20~{\rm keV}$ as an
exclusion range (Fig.~\ref{fig:axsum}).

\subsubsection{Hadronic Axion Window}

This limit as well as the ``trapping side'' of the energy-loss
argument have not been studied in as much detail because the relevant
$m_a$ range is already excluded by the globular-cluster argument
(Fig.~\ref{fig:axsum}) which, however, depends on the axion-photon
interaction which would nearly vanish in models with $E/N=2$. In this
case a narrow gap of allowed axion masses in the neighborhood of 10~eV
may exist between the two SN arguments (``hadronic axion window'').

In this region one can derive interesting limits from globular-cluster
stars where axions can be emitted by nuclear processes, causing a
metallicity-dependent modification of the core mass at helium
ignition~\cite{Haxton91}. It is intruiging that in this window axions
could play a cosmological role as a hot dark matter
component~\cite{Moroi98}. Usually, of course, axions are a {\it
cold\/} dark matter candidate.  Moreover, in this window it may be
possible to detect a $14.4~{\rm keV}$ monochromatic solar axion line
which is produced by transitions between the first excited and ground
state of $^{57}$Fe. In the laboratory one can then search for axion
absorption which would give rise to x-rays as $^{57}$Fe
de-excites~\cite{Moriyama95}.  A recent pilot experiment did not have
enough sensitivity to find axions~\cite{Krcmar98}, but a vastly
improved detector is now in preparation in Tokyo (private
communication by S.~Moriyama and M.~Minowa).

\subsection{Cosmological Limits}
\label{sec:cosmology}

The astrophysical axion mass limits are particularly interesting when
juxtaposed with the cosmological ones which we thus briefly review.
For $f_a\agt 10^8\,{\rm GeV}$ cosmic axions never reach thermal
equilibrium in the early universe.  They are produced by a nonthermal
mechanism that is intimately intertwined with their Nambu-Goldstone
nature and that implies that their contribution to the cosmic density
is proportional to $f_a^{1.175}$ and thus to $m_a^{-1.175}$.  The
requirement not to ``overclose'' the universe with axions thus leads
to a {\it lower\/} mass limit.

One must distinguish between two generic cosmological scenarios.  If
inflation occurred after the Peccei-Quinn symmetry breaking or if
$T_{\rm reheat}<f_a$, the initial axion field takes on a constant
value $a_{\rm i}=f_a \Theta_{\rm i}$ throughout the universe, where
$0\leq\Theta_{\rm i}<\pi$ is the initial ``misalignment'' of the QCD
$\Theta$ parameter~\cite{Preskill83,Abbott83,Dine83,Turner86a}.  If
$\Theta_{\rm i}\sim 1$ one obtains a critical density in axions for
$m_a\sim1~\mu{\rm eV}$, but since $\Theta_{\rm i}$ is unknown there is
no strict cosmological limit on $m_a$.  However, the possibility to
fine-tune $\Theta_{\rm i}$ is limited by inflation-induced quantum
fluctuations which in turn lead to temperature fluctuations of the
cosmic microwave background~\cite{Lyth90,Turner91,Linde91,Shellard98}.
In a broad class of inflationary models one thus finds an upper limit
to $m_a$ where axions could be the dark matter.  According to the most
recent discussion~\cite{Shellard98} it is about $10^{-3}~\rm eV$
(Fig.~\ref{fig:axsum}).

If inflation did not occur at all or if it occurred before the
Peccei-Quinn symmetry breaking with $T_{\rm reheat}>f_a$, cosmic axion
strings form by the Kibble mechanism~\cite{Davis86,Davis89}.  Their
motion is damped primarily by axion emission rather than gravitational
waves.  After axions acquire a mass at the QCD phase transition they
quickly become nonrelativistic and thus form a cold dark matter
component. Unknown initial conditions no longer enter, but details of
the string mechanism are sufficiently complicated to prevent an exact
prediction of the axion density.  On the basis of Battye and
Shellard's treatment~\cite{Battye94a,Battye94b} and assuming that
axions are the cold dark matter of the universe one finds a plausible
mass range of $m_a=\hbox{6--2500$~\mu$eV}$~\cite{Caso98}.  Sikivie
et~al.~\cite{Harari87,Hagmann91,Chang99} predict somewhat fewer
axions, allowing for somewhat smaller masses if axions are the dark
matter. 

Either way, the ongoing full-scale search experiments for
galactic dark matter axions (Sec.~\ref{sec:photonlimits} and
Fig.~\ref{fig:axgam}) in Livermore (U.S.~Axion
Search~\cite{Hagmann98}) and in Kyoto
(CARRACK~\cite{Ogawa96,Yamamoto98}) aim at a cosmologically
well-motivated range of axion masses (Fig.~\ref{fig:axsum}).

%%%%%%%%%%%%%%%%%%%%%%%%%%%%%%%%%%%%%%%%%%%%%%%%%%%%%%%%%%%%%%%%%%%%%%
%% Section 8 %%%%%%%%%%%%%%%%%%%%%%%%%%%%%%%%%%%%%%%%%%%%%%%%%%%%%%%%%
%%%%%%%%%%%%%%%%%%%%%%%%%%%%%%%%%%%%%%%%%%%%%%%%%%%%%%%%%%%%%%%%%%%%%%

\section{LONG-RANGE FORCES}
\label{sec:longrangeforces}

\subsection{Fifth Force}

New low-mass scalar or vector bosons would mediate long-range forces
between macroscopic bodies.  This is in contrast with pseudoscalars
which couple to the spin and thus produce no long-range force between
unpolarized bodies except for a residual force from two-boson
exchange~\cite{Grifols94a,Ferrer98a,Ferrer98b}.  In stars, a new
long-range force has two different consequences.  First, it modifies
the effect of gravity.  Second, it drains the star of energy, for the
quanta of the new force are massless, or nearly so, and thus arise in
thermal reactions.

Thermal graviton emission is a case in
point~\cite{Papini77,Papini89,Schafer83,Gould85,Campo88}.  However,
the graviton luminosity is very small, about $10^{-19}\,L_\odot$ for
the Sun. Naturally, the coherent large-scale force is the most
important aspect of gravity in stars!  This conclusion carries over to
new forces, notably a putative ``fifth force.''  According to
experiment, a fifth force has to be much weaker than
gravity~\cite{Fischbach92,Fischbach96a,Fischbach98,Franklin93} so that
possible modifications of stellar structure~\cite{Glass87,Glass89} or
the solar p-mode frequencies~\cite{Gilliland87,Kuhn88} are too small
to be observable.  Likewise, modifications of fundamental coupling
constants near pulsars~\cite{Ellis89} or scalar boson emission by the
Hulse-Taylor binary pulsar~\cite{Mohanty96} are negligible effects.

However, there are no experimental fifth-force limits below about the
centimeter scale, corresponding to boson masses exceeding about
$10^{-3}~{\rm eV}$, where the most restrictive bounds arise from the
energy loss of stars. The Yukawa coupling $g_S$ ($g_V$) of scalar
(vector) bosons $\phi$ to electrons has been constrained by the
bremsstrahlung process $e^-+{}^4{\rm He}\to{}^4{\rm He}+e^-+\phi$
which leads with Eq.~(\ref{eq:hblimit}) to limits of
$g_S\alt1.3\times10^{-14}$ and $g_V\alt0.9\times10^{-14}$
\cite{Raffelt96a,Grifols86a,Grifols89d}. The Yukawa coupling to
baryons has been constrained by the Compton process $\gamma+{}^4{\rm
He}\to{}^4{\rm He}+\phi$, leading to $g_S\alt4.3\times10^{-11}$ and
$g_V\alt3.0\times10^{-11}$ \cite{Raffelt96a,Grifols86a,Grifols89d}.

\subsection{Leptonic and Baryonic Gauge Interactions}

It has been speculated that lepton and baryon number could play the
role of gauge charges~\cite{Lee55,Okun69}.  One consequence would be
the existence of long-range leptonic and baryonic forces.  The
globular-cluster limits of the previous section translate into
$e_L\alt1\times10^{-14}$ and $e_B\alt3\times10^{-11}$ on the leptonic
and baryonic gauge charges. Tests of the equivalence principle on
solar-system scales constrain a composition-dependent fifth force,
leading to something like $e_{L,B}\alt10^{-23}$~\cite{Blinnikov96}.
The cosmic background neutrinos would screen leptonic forces over
large distances, but the solar-system limit on $e_L$ remains
unaffected~\cite{Dolgov95}.  On the other hand, the SN~1987A neutrino
burst would not suffer dispersion in the leptonic field of the galaxy
because it is shielded by the cosmic background
neutrinos~\cite{Dolgov95}.  Leptonic forces contribute to the neutrino
self-energy, modifying matter-induced neutrino oscillations in the Sun
and supernovae~\cite{Horvat95,Horvat96}.

\subsection{Time-Variation of Newton's Constant}

Astrophysics and cosmology are natural laboratories for testing all
conceivable deviations from the standard theory of
gravitation~\cite{Will93}.  One hypothesis, which goes back to Dirac's
large numbers hypothesis~\cite{Dirac37,Dirac38}, holds that the value
of Newton's constant $G_{\rm N}$ evolves in time. The present-day rate
of change can be measured by a precision study of the orbits of
celestial bodies. In the solar system data come from laser ranging of
the Moon~\cite{Williams96} and radar ranging of the planets, notably
by the Viking landers on Mars~\cite{Shapiro90}. The increase of the
length of day from 1663--1972 caused by tidal forces in the Earth-Moon
system are consistent with a constant $G_{\rm N}$ \cite{Morrison73},
although some controversial claims for a decreasing $G_{\rm N}$ have
been raised~\cite{Will93}.  Beginning in 1974, very precise orbital
data exist for the Hulse-Taylor binary pulsar PSR 1923+16
\cite{Damour91}.  A weaker but less model-dependent bound arises from
the spin-down rate of the pulsar PSR 0655+64~\cite{Goldman90}.
Finally, the long-time stability of galaxy clusters limits a
decreasing $G_{\rm N}$~\cite{Dearborn74}.  The bounds from these
methods are summarized in Table~\ref{tab:gdot}.

\begin{table}[ht]
\centering
\caption{\label{tab:gdot}Range of allowed time variation of
Newton's constant.}
\medskip
\hbox to\hsize{\hss
\begin{tabular}[8]{lrrlll}
\noalign{\hrule\vskip2pt}
\noalign{\hrule\vskip3pt}
Method&\multicolumn{2}{l}{$\dot G_{\rm N}/G_{\rm N}$}&Authors&
Year&Ref.\\
&\multicolumn{2}{l}{$[10^{-12}~{\rm yr}^{-1}]$}\\
&\multicolumn{1}{l}{from}&\multicolumn{1}{l}{to}\\
\noalign{\vskip3pt\hrule\vskip3pt}
Laser ranging (Moon)&$-8$&\hbox{8\quad}&
Williams et al.&1996&\hbox{\cite{Williams96}}\\
Radar ranging (Mars)&$-12$&\hbox{8\quad}
&Shapiro&1990&\cite{Shapiro90}\\
Length of day&$-20$&\hbox{20\quad}
&Morrison&1973&\cite{Morrison73}\\
Binary Pulsar 1913+16&$0$&\hbox{22\quad}
&Damour \& Taylor&1991&\cite{Damour91}\\
Spin-down PSR 0655+64&$-55$&\hbox{55\quad}
&Goldman&1990&\cite{Goldman90}\\
Stability galaxy clusters&$-60$&\hbox{---\quad}
&Dearborn \& Schramm&1974&\cite{Dearborn74}\\
Helioseismology&$-2$&\hbox{2\quad}
&Guenther et al.&1998&\cite{Guenther98}\\
%White-dwarf cooling&---&\hbox{0\quad}
%&Garc\'\i a-Berro et al.&1995&\cite{Garcia95}\\
Globular-cluster ages&$-35$&\hbox{7\quad}
&Degl'Innocenti et al.&1996&\cite{Scilla96}\\
Pulsar masses&$-5$&\hbox{4\quad}
&Thorsett&1996&\cite{Thorsett96}\\
\noalign{\vskip3pt\hrule}
\end{tabular}
\hss}
\end{table}

Very intruiging limits follow from the properties of the Sun.
Paleontological evidence for its past luminosity, which scales
approximately as $G_{\rm N}^7\,{\cal M}^5$, provides limits on
previous values of $G_{\rm N}$~\cite{Teller48}.  Solar models with a
time-varying $G_{\rm N}$ were studied in the 1960s and 1970s
\cite{Gamow67,Pochoda64,Ezer66,Roeder66,Shaviv69,Chin75,Chin76}, but
truly interesting limits arose only recently from helioseismological
observations~\cite{Demarque94,Guenther95,Guenther98}.  A limit $|\dot
G_{\rm N}/G_{\rm N}|\alt 2\times10^{-12}~{\rm yr}^{-1}$ was derived by
a comparison of the measured small-spacing p-mode frequency
differences with those calculated from solar models with a
time-varying $G_{\rm N}$~\cite{Guenther98}. The authors believe that
the uncertainty is dominated by the observational errors while the
prime systematic uncertainty is the exact solar age.  A more
conservative approach was used in deriving limits on a new solar
energy loss mechanism (Sec.~\ref{sec:helioseismology}).  The
helioseismological analysis of~\cite{Guenther98} provides the most
restrictive limit on $\dot G_{\rm N}/G_{\rm N}$, hopefully stimulating
other groups to re-assess the bound independently.

A large effect is expected on the oldest stars which ``integrate''
$G_{\rm N}(t)$ into the distant past.  White-dwarf cooling is a case
in point~\cite{Vila76,Garcia95}.  Under reasonable assumptions for the
galactic age, the faint end of the luminosity function prefers a
negative value for $\dot G_{\rm N}/G_{\rm N}$ around $-10$ to
$-30\times10^{-12}~{\rm yr}^{-1}$ \cite{Garcia95}. Very recently this
case has been re-examined~\cite{Benvenuto99} in greater detail. The
observational uncertainty of the faint end of the luminosity function
and the uncertainty of the galactic age preclude a clear limit on
$\dot G_{\rm N}/G_{\rm N}$. However, it is remarkable that even values
as small as $10^{-14}~{\rm yr}^{-1}$ seem to make a noticeable
difference for the cooling behavior of the oldest white dwarfs.

Globular clusters are another important case because a different
$G_{\rm N}$ in the past changes their apparent age based on the
brightness of the main-sequence turn-off
\cite{Prather76,VandenBerg77,Scilla96}.  A comparison with the
expansion age of the universe brackets the allowed rate-of-change to
the interval shown in Table~\ref{tab:gdot}.

A very sensitive limit arises from the observed masses of several old
pulsars which measure the value of $G_{\rm N}$ at their time of
formation in a SN explosion~\cite{Thorsett96}.  The mass of a SN core
at the time of collapse depends on its Chandrasekhar value which in
turn scales as ${G_{\rm N}}^{-3/2}$.

The limits shown in Table~\ref{tab:gdot} are difficult to compare on
an equal footing as they involve vastly different ways of dealing with
statistical and systematic uncertainties.  However, it looks fair to
conclude that $|\dot G_{\rm N}/G_{\rm N}|$ cannot exceed a few
$10^{-12}~{\rm yr}^{-1}$ and that stars play an important role in this
discourse.  A similar bound arises from the cosmic expansion rate at
the time of big-bang nucleosynthesis as measured by the primordial
light-element abundances~\cite{Malaney93}.  It implies that three
minutes after the big bang $G_{\rm N}$ agreed with its present-day
value to within a few tens of percent.  A comparison of this limit
with those of Table~\ref{tab:gdot} requires a specific assumption
about the functional dependence of $G_{\rm N}(t)$.

\subsection{Equivalence Principle}

The general relativistic equivalence principle implies that the
space-time trajectories of relativistic particles are independent of
internal degrees of freedom such as spin or flavor, and independent of
the particle type (e.g.\ photon, neutrino). Several astronomical
observations allow tests of this prediction.

Limits on a gravitationally induced birefringence effect for photon
propagation have been derived from the absence of depolarization of
the Zeeman components of spectral lines emitted in magnetically active
regions of the Sun~\cite{Gabriel91a}. Observations of the light
deflection by the Sun could soon become interesting~\cite{Gabriel91b}.
The depolarization effect on distant radio galaxies already provides
very restrictive limits~\cite{Carroll91,Haugan95}, as do pulsar
observations~\cite{LoSecco89,Klein90,Krisher91}.

One may also test for the equality of the Shapiro time delay between
different particles which propagate through the same gravitational
field.  The absence of an anomalous shift between the SN~1987A photon
and neutrino arrival times (Sec.~\ref{sec:signaldispersion}) gave
limits on violations of the equivalence
principle~\cite{Krauss88,Coley88,Almeida89}.  The observation of a
future galactic SN could provide independent arrival information for
$\bar\nu_e$ and $\nu_e$ and thus provide another such
test~\cite{Pakvasa89,LoSecco88}.

A violation of the equivalence principle could manifest itself by a
relative shift of the energies of different neutrino flavors in a
gravitational field. For a given momentum $p$ the matrix of energies
in flavor space (relativistic limit) is $E=p+M^2/2p+2p\phi({\bf
r})(1+F)$ where $M^2$ is the squared neutrino mass matrix, $\phi({\bf
r})$ the Newtonian gravitational potential, and $F$ a matrix of
dimensionless constants which parametrize the violation of the
equivalence principle.  $F\not=0$ can lead to neutrino oscillations in
analogy to the standard vacuum oscillations which are caused by the
matrix $M^2$
\cite{Gasperini88,Gasperini89,Halprin91,Pantaleone93,Iida93,%
Minakata95,Bahcall95a,Halprin96,Mureika97,Glashow97,Halprin98,%
Mansour98,Casini98}.
Values for $F_{ij}$ in the general $10^{-14}{-}10^{-17}$ range could
account for the solar neutrino problem.

\subsection{Photon Mass}

While it is usually taken for granted that photons are strictly
massless, this theoretical expectation still needs to be tested
experimentally. Some of the most restrictive constraints are related
to the long-range nature of static electric or magnetic fields.  The
best laboratory limit of $m_\gamma\alt10^{-14}~{\rm eV}$ derives from
a test of Coulomb's law---see~\cite{Goldhaber71} for a review.

In the astrophysical domain, the dispersion of the pulsed signal of
radio pulsars is not a very sensitive diagnostic as the interstellar
medium mimics a photon mass corresponding to a plasma frequency of
order $10^{-11}~{\rm eV}$.  The spatial variation of magnetic fields
of celestial bodies is far more sensitive.  Jupiter's magnetic field
as measured by Pioneer-10 yields $m_\gamma\alt0.6\times10^{-15}~{\rm
eV}$~\cite{Davis75} while the Earth's field gives
$m_\gamma\alt0.8\times10^{-15}~{\rm eV}$~\cite{Fischbach94}.

If the photon has a mass, $A\,m_\gamma^2$ is an observable quantity
where $A$ is the vector potential corresponding to known magnetic
fields. A recent laboratory experiment discloses
$A\,m_\gamma^2\alt0.8\times10^{-22}~{\rm T~m~eV^2}$ \cite{Lakes98}.
The galactic magnetic field implies $A\approx 2\times10^9~{\rm T~m}$
(Tesla-meter) so that $m_\gamma\alt2\times10^{-16}~{\rm eV}$ while a
cluster-level field corresponds to $A\approx10^{12}~{\rm T~m}$,
providing $m_\gamma\alt10^{-17}~{\rm eV}$.

Even more restrictive limits obtain from astrophysical objects in
which magnetic fields, and hence the Maxwellian form of
electrodynamics, play a key role at maintaining equilibrium or
creating long-lived stable structures~\cite{Barrow84}. The most
restrictive case is based on an argument about the
magneto-gravitational equilibrium of the gas in the Small Magellanic
Cloud. The argument requires that the range of the interaction exceeds
the characteristic field scale of about 3~kpc~\cite{Chibisov76}.  This
resulting limit $m_\gamma\alt10^{-27}~{\rm eV}$, if correct, is
surprisingly close to $10^{-33}~{\rm eV}$ where the photon Compton
wavelength would exceed the radius of the observable universe and thus
would cease to have any observable consequences.

\subsection{Multibody Neutrino Exchange}

Two-neutrino exchange between fermions gives rise to a long-range
force.  A neutrino may also pass around several fermions, so to speak,
producing a much smaller potential.  In a thought-provoking paper it
was claimed that this multibody neutrino exchange could be a huge
effect in neutron stars, essentially because combinatorial factors
among many neutrons win out against the smallness of the
potential~\cite{Fischbach96b}.  To stabilize neutron stars, it was
claimed, the long-range nature of neutrino exchange had to be
suppressed by a nonvanishing mass exceeding about $0.4~{\rm eV}$ for
all flavors. In an interesting series of papers it was shown, however,
that a proper resummation of a seemingly divergent series of terms
leads to a well-behaved and small ``neutron-star self-energy''
\cite{Kiers98,Abada96,Abada98a,Abada98b,Abada98c,Arafune98},
invalidating the claim of a lower neutrino mass limit.

%%%%%%%%%%%%%%%%%%%%%%%%%%%%%%%%%%%%%%%%%%%%%%%%%%%%%%%%%%%%%%%%%%%%%%
%% Conclusion %%%%%%%%%%%%%%%%%%%%%%%%%%%%%%%%%%%%%%%%%%%%%%%%%%%%%%%%
%%%%%%%%%%%%%%%%%%%%%%%%%%%%%%%%%%%%%%%%%%%%%%%%%%%%%%%%%%%%%%%%%%%%%%

\section{CONCLUSION}
\label{sec:conclusion}

Stellar-evolution theory together with astronomical
observations, the SN~1987A neutrino burst, and certain x- and
$\gamma$-ray observations provide a number of well-developed arguments
to constrain the properties of low-mass particles.  The most
successful examples are globular-cluster stars where the ``energy-loss
argument'' was condensed into the simple criteria of
Eqs.~(\ref{eq:rglimit}) and~(\ref{eq:hblimit}) and SN~1987A where it
was summarized by Eq.~(\ref{eq:snlimit}).  New particle-physics
conjectures must first pass these and other simple astrophysical
standard tests before being taken too seriously.

A showcase example for the interplay between astrophysical limits with
laboratory experiments and cosmological arguments is provided by the
axion hypothesis.  The laboratory and astrophysical limits push the
Peccei-Quinn scale to such high values that it appears almost
inevitable that axions, if they exist at all, play an important role
as a cold dark matter component.  This makes the direct search for
galactic axion dark matter a well-motivated effort.  Other important
standard limits pertain to neutrino electromagnetic form
factors---laboratory experiments will have a difficult time catching
up.

The globular-cluster limit was based on relatively old observational
data. A plot like Fig.~\ref{fig:coremass} could be made more
significant with dedicated CCD observations of globular clusters and
improved theoretical interpretations. Assuming that such an effort
produces internally consistent results, the statistical significance
would improve, but I would not expect a vast gain for, say, the
neutrino magnetic-moment limit as there always remain irreducible
systematic uncertainties.

Shockingly, SN~1987A as a particle-physics laboratory is based on no
more than two dozen measured neutrinos. The observation of a future
galactic SN with a large detector like Superkamiokande or a future
observatory such as OMNIS~\cite{Smith97} would provide a
high-statistics neutrino light curve and thus a sound empirical basis
for SN theory in general and for particle-physics interests in
particular.  Alas, galactic supernovae happen only once every few
decades, perhaps only once per century. Thus, while the neutrinos from
the next galactic SN surely are on their way, it could be a long wait
until they arrive.

Most of the theoretical background relevant to this field could not be
touched upon in this brief overview. The physics of weakly coupled
particles in stars is a nice playing field for ``particle physics in
media'' which involves field theory at finite temperature and density
(FTD), many-body effects, particle dispersion and reactions in
magnetic fields and media, oscillations of trapped neutrinos, and so
forth. It is naturally in the context of SN theory where such issues
are of particular interest, but even the plasmon decay
$\gamma\to\nu\bar\nu$ in normal stars or the MSW effect in the Sun are
interesting cases. Particle physics in media and its astrophysical and
cosmological applications is a fascinating topic in its own right
which well deserves a dedicated review.

Much more information of particle-physics interest may be written in
the sky than has been deciphered as yet.  Other objects or phenomena
should be considered, perhaps other kinds of conventional stars,
perhaps more exotic phenomena such as $\gamma$-ray bursts.  The
particle-physics lessons to be learned from them are left to be
reviewed in a future report!

\section*{ACKNOWLEDGMENTS}

This work was supported, in part, by the Deutsche
Forschungsgemeinschaft under grant No.\ SFB-375.

%%%%%%%%%%%%%%%%%%%%%%%%%%%%%%%%%%%%%%%%%%%%%%%%%%%%%%%%%%%%%%%%%%%%%%
%% References %%%%%%%%%%%%%%%%%%%%%%%%%%%%%%%%%%%%%%%%%%%%%%%%%%%%%%%%
%%%%%%%%%%%%%%%%%%%%%%%%%%%%%%%%%%%%%%%%%%%%%%%%%%%%%%%%%%%%%%%%%%%%%%

\end{document}